\documentclass[aps,
        twocolumn,
        superscriptaddress,
        10pt]{revtex4-1}

\usepackage{color}
\usepackage{amsmath}
\usepackage{graphicx}
\usepackage{epstopdf}
\usepackage{dcolumn}
\usepackage{bm}
\usepackage{textcomp}
\usepackage[font=small,format=plain,justification=justified]{caption}
\usepackage{subcaption}
\usepackage{gensymb}

\newcommand{\bE}{\mathbf{E}}
\newcommand{\bH}{\mathbf{H}}
\newcommand{\bJ}{\mathbf{J}}
\newcommand{\sw}{\text{sw}}
\newcommand{\bX}{\mathbf{X}}

\newcommand{\xhat}{\hat{x}}
\newcommand{\yhat}{\hat{y}}
\newcommand{\zhat}{\hat{z}}
\newcommand{\tH}{\tilde{H}}

\renewcommand{\t}[1]{\text{#1}}

\def\=#1{\overline{\overline #1}} 

\begin{document}


\title{Near-Perfect
Conversion of a Propagating Plane Wave into a Surface Wave
Using Metasurfaces}
\author{S.~N.~Tcvetkova}
\email{svetlana.tcvetkova@aalto.fi}
\affiliation{Department of Electronics and Nanoengineering\\
Aalto University, P.O. Box 15500, 00076 Aalto, Finland}
\author{D.-H. Kwon}
\affiliation{Department of Electrical and Computer Engineering\\
University of Massachusetts Amherst, Amherst, Massachusetts 01003, USA}
\author{A. D\'{i}az-Rubio}
\affiliation{Department of Electronics and Nanoengineering\\
Aalto University, P.O. Box 15500, 00076 Aalto, Finland}
\author{S.~A.~Tretyakov}
\affiliation{Department of Electronics and Nanoengineering\\
Aalto University, P.O. Box 15500, 00076 Aalto, Finland}

\date{\today}

\begin{abstract}
In this paper, theoretical and numerical studies
of perfect/nearly-perfect conversion of a plane wave into 
a surface wave are presented. 
The problem of determining the electromagnetic properties of 
an inhomogeneous lossless boundary which would fully transform 
an incident plane wave into a surface wave propagating along 
the boundary is considered. An approximate field
solution which produces a slowly growing surface wave and
satisfies the energy conservation law is discussed and 
numerically demonstrated. The results of the study 
are of great importance for the future development of 
such devices as perfect leaky-wave antennas and can potentially
lead to many novel applications.
\end{abstract}

\maketitle


\section{\label{sec:1}Introduction}

Traditional leaky-wave antennas~\cite{oliner2007} 
at microwave frequencies are devices that
convert between space and guided wave propagation modes.
By introducing either periodic or continuous perturbation to {\color{black} a} waveguide or transmission-line structure, portion of
the guided power is designed to leak in a desired radiation direction.
For a standard leaky-wave structure,
the complex propagation constant describes the rate of exponential
amplitude decay and the phase velocity along the guided wave direction.
The characteristics of the radiated wave can be obtained from the
field distribution over
the radiating aperture. A comprehensive
review on leaky-wave theory and techniques is available in
\cite{monticone_procieee2015}.

Since the aperture field distribution of a standard leaky-wave
structure is different from a
uniform-amplitude and linear-phase one,
the conversion between the space and guided waves is not perfect.
In the transmitting case, this translates to radiation into unwanted
directions; in the receiving case, it means that the incoming
plane wave from the scan direction is partially reflected
and scattered, rather than completely transformed into
the guided-mode wave. Recently, synthesis of a desired leaky-wave
radiation characteristics associated with a custom aperture field
distribution is receiving an increased interest using
spatially varying perturbation
structures---the waveguide width and the metallic post interval
in a substrate-integrated waveguide~\cite{martinez-ros_ieeejap2012};
tensor sheet admittances in stacked metasurfaces on a ground
plane~\cite{tierney_ieeeaps2015}; and
the shape, size, and periodicity of locally anisotropic unit cell
in the form of printed conductor patches on a grounded dielectric
substrate~\cite{minatti_ieeejap2016b,minatti_ieeejap2016c}.

The problem of conversion of a propagating plane wave into 
a surface wave appears to be similar to that
of anomalous reflection or refraction in underlying physics, 
where a propagating plane wave is converted into another 
\emph{propagating} plane wave. 
Recently, it was recognized that manipulation of propagating
waves using such devices as conventional reflectarrays and 
transmitarrays is accompanied by fundamental imperfections, associated with
inevitable power scattered into undesired directions.
Studies of non-local metasurfaces have demonstrated
the possibility to create devices, which can perfectly transform 
a plane wave incident along one direction into a plane 
wave propagating into another direction
\cite{asadchy_prb2016,estakhri_prx2016}.
Now, both theoretical designs employing penetrable and impenetrable
metasurfaces~\cite{epstein_prl2016,kwon_prb2017} 
and a practical realization
based on the leaky-wave principle on a super-cell
level~\cite{diaz-rubio_sciadv2017} are available.
However, in sharp contrast to anomalous reflection, perfect conversion between space waves and guided waves
is still elusive. No designs are available for the canonical
transformation problem between a plane wave and a surface wave
that promise perfect conversion devoid of spurious scattering
in the limit of lossless constituents.

In \cite{sun_natmater2012},
the authors reported a periodic metasurface with linear reflection phase 
approximation based on the generalized law of
reflection~\cite{yu_science2011} (similarly to conventional gratings),
in which a power conversion efficiency  of nearly a 100\%
was claimed.
However, the reported structure cannot operate 
without introducing losses
{\color{black}for an infinitely long structure},
because otherwise a momentum mismatch between 
a propagating wave and a surface wave appears, not being able
to excite a surface wave (full reflection of the propagating 
wave back into the free space). 
Furthermore, such a structure does not support a surface wave to propagate 
along the surface as the surface wave is not an eigensolution
of the metasurface. As a result, the reported gradient metasurface,
{\color{black}as an infinitely long converter},
operates as a good absorber at steady state.
{\color{black}For finite-length converters, the power conversion
efficiency between a plane wave and a surface wave
was theoretically and numerically studied in
\cite{qu_epl2013}. For periodic supercell-based gradient
metasurface implementations, the decoupling effect taking
place at the interface between supercells has a significant
impact on efficiency. For a converter comprising two supercells,
a high conversion efficiency of 78\% was predicted.
}

Further, the authors also developed a
surface plasmon polaritons (SPP) approach {\cite{sun_lsa2016}} based on the generalized law of reflection~\cite{yu_science2011}. The model consists of a
meta-coupler, which is impedance matched to the free space 
and plasmonic metal sheet located below. Such a system allows 
creation of an SPP wave without decoupling effect. The authors 
claim to have a numerically predicted conversion efficiency of $94\%$.
The design consists of two surfaces with an
approximately half-wavelength separation between them.
A single, low-profile design would have been more desirable.
Furthermore, an SPP generated by a meta-coupler upon plane-wave
illumination will go through additional refraction by the same
meta-coupler with the wavenumber along the surface shifted further into
the invisible range. This effect of higher-order spatial
harmonics generation deep in the invisible range has not been
investigated.

In \cite{estakhri_prx2016}, transformation
of a propagating wave into a surface wave using metasurface
was described, which is designed as a passive and lossy 
periodic structure. However, the reported 
conversion efficiency of such a metasurface is quite low ($\approx 7\%$).
A transparent metasurface for transforming a beam wave into a 
surface wave and back into a beam wave was reported in
\cite{achouri_metamaterials2016, achouri_arxiv2016}. Contrary to the claims,
the propagating beam launched by the metasurface appears to be
generated by active (i.e., source) metasurface constituents rather than being converted from the surface wave propagating along the metasurface.

In this paper, we theoretically and numerically investigate conversion of propagating plane waves to surface waves. In particular, we explain that it is not possible to create a {\color{black} point-wise lossless}  metasurface which would perfectly  convert an incident plane wave into a single surface-wave mode carrying linearly growing power along the propagation direction (as required by the energy conservation). On the other hand, we present an approximate solution for the surface impedance of a metasurface which performs 
such a conversion with a high (predicted to be nearly 100\%) efficiency. As an example, near-perfect conversion of a plane wave into a surface wave with a slow exponential growth is numerically tested.
Such a solution is a special case of separable solutions to the Helmholtz equation, which consist of a single spatial harmonic. Although in this case the power growth law is different from the ideal linear dependence, we show that a very accurate approximate solution can be found if the exponential growth of a surface-bound eigenwave is slow enough.
The appropriate surface impedance that realizes the envisioned conversion is found from the boundary condition using the total field distribution.
Example designs are numerically tested to demonstrate near-perfect
propagating wave-to-surface wave conversion performance. {\color{black} Furthermore, we show that non-local metasurfaces can emulate the active-lossy behaviour necessary for ideal conversion, without the need to use any active or dissipating components.}  

In the microwave regime, the plane wave-to-surface wave
converting surface in this study may be realized as a thin metasurface
on an impenetrable surface, e.g., as an array of sub-wavelength
resonators printed on a grounded dielectric substrate.
Such a single-surface design overcomes the SPP-to-space wave
decoupling issue associated with existing meta-coupler designs
while maintaining a low profile.

\section{Problem statement}
\label{sec:problem}

\begin{figure}[t]
	\centering
		\includegraphics[width=0.45\textwidth]{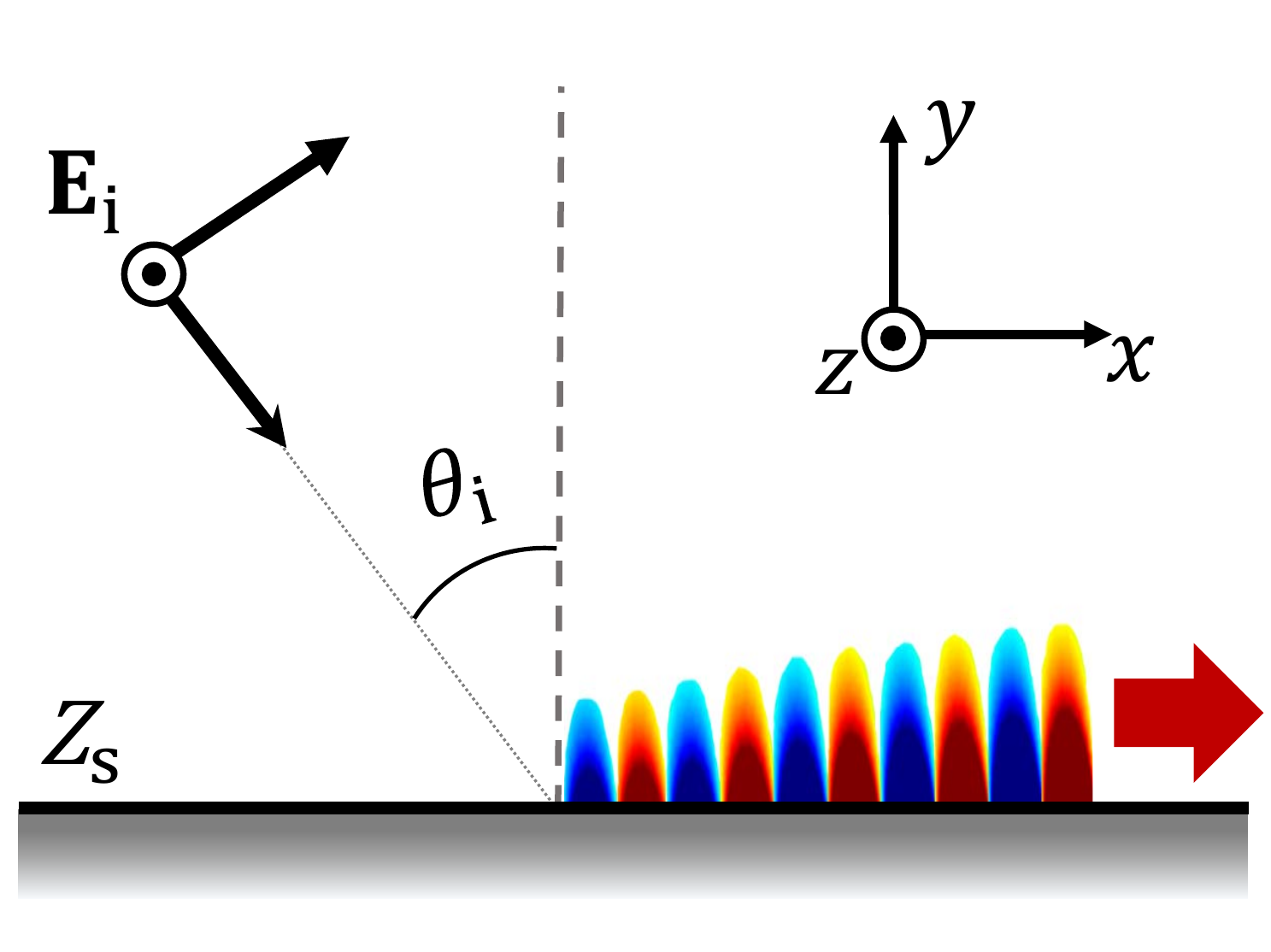}
	\caption{Boundary problem of plane wave to surface wave conversion.}
	\label{fig:1}
\end{figure}

Figure~\ref{fig:1} illustrates the problem under consideration,
perfect conversion {\color{black}(no dissipation or
scattering losses)} of a TE-polarized incident plane wave into a TM-polarized surface wave bound to an impenetrable metasurface in the $xz$-plane, characterized by the matrix surface impedance $\=Z_{\rm s}$. 
Polarization transformation is applied here to avoid field and power interference of propagating and surface waves with each other~\cite{asadchy_prb2016}. 
Both fields are invariant with respect to $z$.
The design objective is to
find $\=Z_{\rm s}$ that enables this transformation. 
{\color{black} The most desirable solution is a point-wise lossless metasurface, such that the normal component of the total Poynting vector is zero at all points of the surface. In this case, the impedance matrix is skew-Hermitian:
$\=Z_\text{s}^\dag=-\=Z_\text{s}$.
In the special case of reciprocal surfaces, all components of the surface impedance matrix of lossless boundaries are purely imaginary \cite{pozar2005}.}

\begin{figure}[h]
	\centering
		\includegraphics[width=0.45\textwidth]{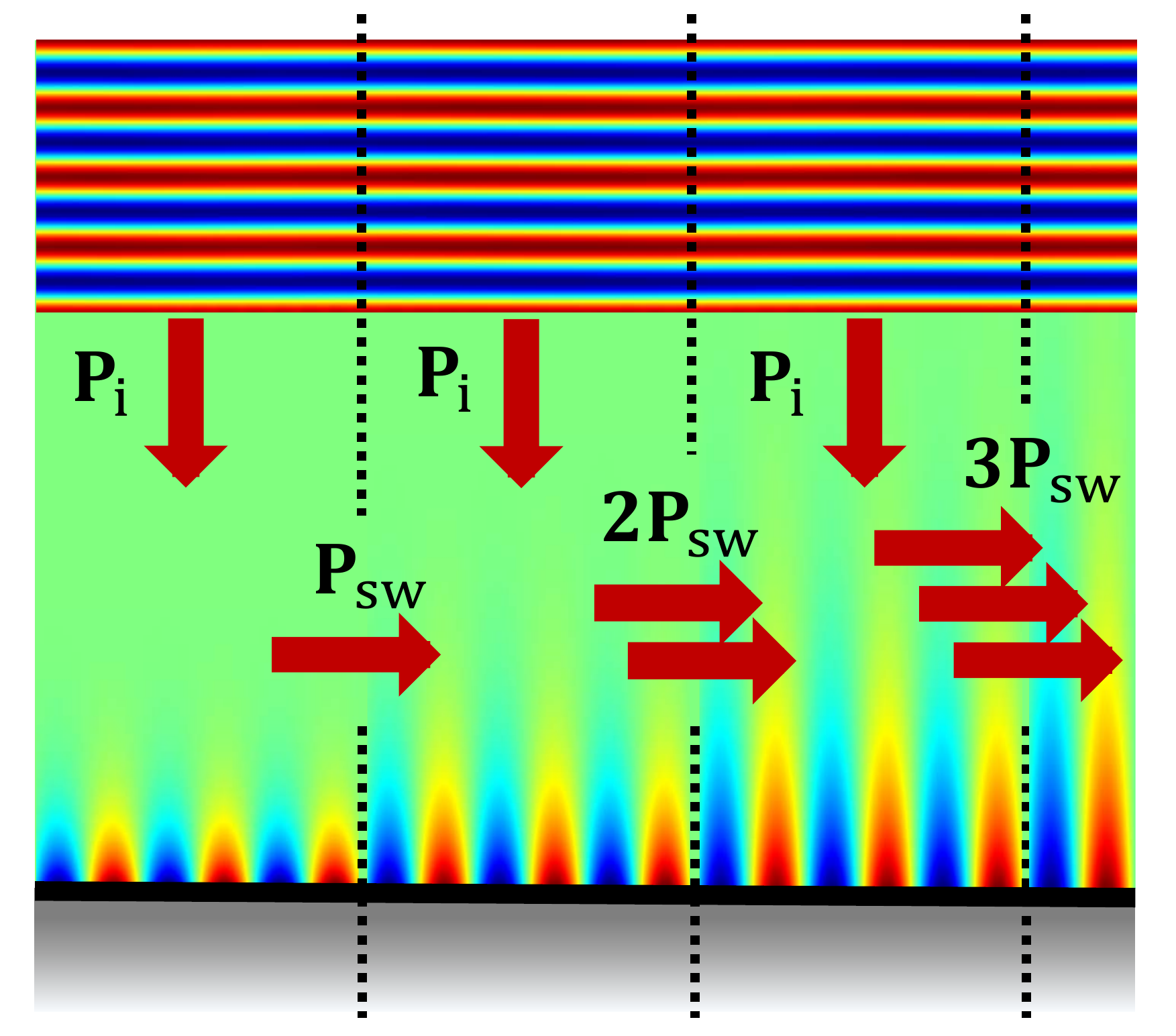}
			\caption{{\color{black}Illustration of a linear power growth required
by the energy conservation law.}}
	\label{fig:power}
\end{figure}

Both the incident fields and the scattered surface-wave fields must
satisfy Maxwell's equations in the upper half-space, which is
assumed to be free space. 
To fully understand the concept of energy transfer and its further flow along the surface, a point-wise lossless discretized surface should be considered (Figure~{\ref{fig:power}}). The uniform amount of power carried by the incident wave  (denoted as ${\bf P}_{\rm i}$ in Fig.~{\ref{fig:power}}) is added at each consecutive small interval. Therefore, the power carried by the surface wave should grow linearly along the surface, to satisfy the energy conservation.

In search of a surface-wave solution that satisfies
the necessary spatial power dependence, both separable and
non-separable solutions to the Helmholtz equation have been considered.
Conventional separable solutions are plane waves which exponentially grow along the surface.
Using an $e^{j\omega t}$ time convention assumed and suppressed
for time-harmonic analysis at an angular frequency $\omega$,
the non-separable solutions in Cartesian
coordinates~\cite{moseley_qam1965,schoonaert_eleclett1973}
allow the $z$-component of the magnetic field in TM polarization
of the form
\begin{equation}
H_z=H_0(k_yx-k_xy)e^{-j(k_xx+k_yy)},
\label{Hz:nonseparable}
\end{equation}
where $H_0$ is a constant.
The wavenumbers in the $x$- and $y$-directions, $k_x$ and $k_y$,
satisfy the free-space dispersion relation:
\begin{equation}
k_x^2+k_y^2=k_0^2,
\label{fs_dispersion}
\end{equation}
where $k_0$ is the free-space wavenumber.
An evanescent wave in the $y$-direction is associated with the
condition $k_x>k_0$. It is found that the time-average power carried
in the $x$-direction increases quadratically with $x$. Hence, both separable and the non-separable (\ref{Hz:nonseparable}) eigenwave field solutions cannot represent the surface wave converted from the incident plane wave.
 In other words, Maxwell's equations in homogeneous media do not allow ideal conversion of uniform propagating plane waves into one surface wave with the use of point-wise lossless metasurfaces.
Therefore, we introduce approximate nearly perfect solutions and explore possibilities offered by active/lossy and strongly non-local metasurfaces.

\section{Separable field solutions}
\label{sec:separable}

Let us consider a separable solution for the surface wave, whose magnetic field
is expressed as
\begin{equation}
\bH^\sw=\zhat H^\sw_z(x,y)=\zhat X(x)Y(y),
\label{Hsw_separable}
\end{equation}
where $X(x)$ and $Y(y)$ are functions of $x$ and $y$ only,
respectively. The surface wave field component satisfies the
homogeneous wave equation, $(\nabla^2+k_0^2)H^\sw_z=0$.
Following the standard separation-of-variables
technique~\cite{harrington2001}, one obtains a solution
of the form
\begin{equation}
H^\sw_z=H^\sw_0e^{-j(k'_xx+k'_yy)}
\label{Hswz:separable}
\end{equation}
with an arbitrary complex amplitude $H^\sw_0$. The wavenumbers
$k'_x$, $k'_y$ satisfy (\ref{fs_dispersion}). The prime symbol denotes spectral variables.

Let us write the wavenumbers in terms of real-valued
propagation ($\beta$'s) and attenuation ($\alpha$'s) constants as
$k'_x=\beta'_x-j\alpha'_x$, $k'_y=\beta'_y-j\alpha'_y$. {\color{black} Because perfect transformation into a single surface mode using point-wise lossless surfaces is not possible, we consider 
the general
solution for $H^\sw_z$ in $y\geq 0$ as} a superposition of
(\ref{Hswz:separable}) over the entire complex-$k_x$ plane, written
as a two-dimensional (2-D) inverse Fourier transform 
\begin{equation}
H^\sw_z=\frac{1}{4\pi^2}
\iint\limits_{-\infty}^\infty\tH^\sw_z(k'_x,k'_y)
e^{-j(k'_xx+k'_yy)}d\alpha'_xd\beta'_x,
\label{Hswz:2d_fourier}
\end{equation}
where $\tH^\sw_z(k'_x,k'_y)$ is the 2-D spectrum of $H^\sw_z(x,y)$.
{\color{black}In (\ref{Hswz:2d_fourier}), the complex propagation
constant $k'_y$ is found from the free-space dispersion relation
(\ref{fs_dispersion}). The branch of the square root
for $\alpha'_y$ is determined such that
$\beta'_y\geq 0$, i.e., all scattered wave components  propagate
away from the $y=0$ boundary.}
Equation~(\ref{Hswz:2d_fourier}) represents superposition of
homogeneous and inhomogeneous plane waves. For (\ref{Hswz:2d_fourier})
to represent a surface wave bound to the $xz$-plane,
$\tH^\sw_z(k'_x,k'_y)$ can be non-zero only in the range $\alpha'_y>0$.
If we desire that the converted surface wave propagates in the
$+x$-axis direction along the surface, the valid region of
non-zero $\tH^\sw_z(k'_x,k'_y)$ in the complex-$k_x$ plane corresponds
to $\beta'_x>k_0$.

Using Maxwell's equations, expressions for
the E-field components of the surface
wave, $\bE^\sw=\xhat E^\sw_x+\yhat E^\sw_y$, are found to be
\begin{eqnarray}
E^\sw_x &=& -\frac{1}{4\pi^2}\iint\limits_{-\infty}^\infty
\frac{\eta_0k'_y}{k_0}\tH^\sw_ze^{-j(k'_xx+k'_yy)}d\alpha'_xd\beta'_x,
\label{Eswx:2d_fourier}\\
E^\sw_y &=& \frac{1}{4\pi^2}\iint\limits_{-\infty}^\infty
\frac{\eta_0k'_x}{k_0}\tH^\sw_ze^{-j(k'_xx+k'_yy)}d\alpha'_xd\beta'_x,
\label{Eswy:2d_fourier}
\end{eqnarray}
where $\eta_0\approx 377~\Omega$ is the free-space intrinsic impedance.
With the general expression of the surface wave fields in
(\ref{Hswz:2d_fourier})--(\ref{Eswy:2d_fourier}), the surface wave
design reduces to finding $\tH^\sw_z$ that performs the
following functions:
1) elimination of a reflected plane wave from the surface and
2) linear growth of the surface-wave power with respect to $x$
that is consistent with perfect conversion from the incident plane wave.

\section{Approximate design employing a single surface wave of slow exponential growth}
\label{sec:slow_exp_growth}

It is expected that there are many possibilities
for the spectrum $\tH^\sw_z(k_x',k_y')$ that satisfy the design
requirements.
In order to reduce the design complexity, let us consider
the special case of a single spatial harmonic,
such that the surface wave spectrum is represented
as
\begin{equation}
\tH^\sw_z(k_x',k_y')=4\pi^2H^\sw_0\delta(k_x'-k_x)\delta(k_y'-k_y), 
\label{def:sp_hsw}
\end{equation}
where $H^\sw_0$ is a complex amplitude and
$k_x=\beta_x-j\alpha_x$, $k_y=\beta_y-j\alpha_y$ are
complex propagation constants of choice in the $x$- and 
$y$-axis directions, respectively. The associated 
surface-wave field expressions are given by
\begin{eqnarray}
\bH^\sw &=&\zhat H^\sw_0 e^{-(\alpha_x+j\beta_x)x}e^{-(\alpha_y+j\beta_y)y},
\label{def:hsw}\\
\bE^\sw &=&
\left[-\xhat\left(\beta_y-j\alpha_y\right)
+\yhat\left(\beta_x-j\alpha_x\right)\right]\frac{\eta_0H^\sw_0}{k_0}
\nonumber\\
&&\times e^{-(\alpha_x+j\beta_x)x}e^{-(\alpha_y+j\beta_y)y}.
\label{def:esw}
\end{eqnarray}
Obviously, the power carried by  this simple single-mode surface wave grows exponentially along $x$, while the ideal conversion of an incident propagating plane wave into this surface wave implies linear power increase. However, solutions with a slow exponential growth can approximate the required linear law.

For complex propagation constants $k_x$ and $k_y$,
the free-space dispersion relation (\ref{fs_dispersion})
gives two separate equations for real-valued parameters
$\alpha_x$, $\beta_x$, $\alpha_y$, and $\beta_y$ defined by
\begin{gather}
\beta_x^2+\beta_y^2-\alpha_x^2-\alpha_y^2=k_0^2,
\label{fs_disp_real}\\
\alpha_x\beta_x+\alpha_y\beta_y=0.
\label{fs_disp_imag}
\end{gather}
Any combination of real values for the four parameters that satisfy
(\ref{fs_disp_real})--(\ref{fs_disp_imag}) produce fields that satisfy
Maxwell's equations in the range $y>0$ in Fig.~\ref{fig:1}.
Here, we choose
$\alpha_x$ to be 
a small negative quantity and $\beta_x>k_0$.
This results in $\alpha_y>0$, $0<\beta_y<k_0$.
Such a wave represents a propagating wave of slow exponential
growth in the $+x$-axis direction. Yet, the wave does not reach
$y\to +\infty$, bound to the $xz$-plane, owing to $\alpha_y>0$.

For realization of the wave converter as a point-wise lossless non-transparent
metasurface, the resulting total fields, given as a superposition
of the incident plane wave and the slowly-growing surface wave,
must have zero net power across the $xz$-plane
everywhere~\cite{kwon_prb2017}. This condition is analogous
to the local power conservation requirement for passive, lossless
realization of $\Omega$-bianisotropic metasurfaces for
wave transformation~\cite{epstein_prl2016}.
The total power density 
along the normal to the surface may 
be dependent on the vertical coordinate and be inhomogeneous, 
but we require it to be equal zero on the lossless metasurface boundary  ($y=0$).
This assures that all the illuminating power of the incident wave is ``accepted''
by the surface and consequently used for surface wave creation. 
For simplicity, let us consider a
normally incident plane wave in Fig.~\ref{fig:1} with the fields
given by
\begin{equation}
\bE^i=\zhat E^i_0 e^{jk_0y},\ \ 
\bH^i=-\xhat\frac{E^i_0}{\eta_0}e^{jk_0y}.
\label{def:ei_hi}
\end{equation}
On the surface $(y=0)$, the zero net power penetration condition reads
\begin{equation}
\begin{split}
S_{y}(x,y=0) &= S^{\rm i}_{ y} + S^{\rm sw}_{ y}\\
& = - {{|E_0|^2}\over {2 \eta_0}}+{{\eta_0 \beta_{ y}}\over {2 k_0} } |H^{\rm sw}_0|^2 \, e^{-2 \alpha_{ x} x}=0,
\end{split}
\label{cond_sy}
\end{equation}
where $S^{\rm i}_y$ and $S^\sw_y$ 
are the normal components of the time-average Poynting vector
associated with the incident plane wave and the surface wave,
respectively. The normal component of the Poynting vector for the
total fields, $S_y$, is an algebraic sum because the two sets of fields
are of orthogonal polarizations. Hence, for the surface wave
fields (\ref{def:hsw})--(\ref{def:esw}) to cancel the incident power
density on the surface, the magnetic field magnitude of the surface
wave must satisfy
\begin{equation}
|H^\sw_0|=\frac{|E_0|}{\eta_0}\sqrt{\frac{k_0}{\beta_y}}e^{\alpha_x x}.
\label{cond:hsw0mag}
\end{equation}
The phase of $H^\sw_0$ relative to $E_0$ can be arbitrary.
However, if $H^\sw_0$ satisfies (\ref{cond:hsw0mag}),  the surface wave amplitude does not grow along the surface, and, moreover, 
the complex amplitude should be a constant to satisfy Maxwell's equation.
Therefore, it is concluded that (\ref{cond_sy}) cannot
be satisfied for all $x$. In fact, (\ref{cond_sy}) can be exactly satisfied
only at one location in $x$ because $\alpha_x\neq 0$~\footnote{If
$\alpha_x=0$, (\ref{fs_disp_imag}) gives
two possibilities: $\alpha_y=0$ or $\beta_y=0$. The former results
in a propagating plane wave rather than a surface wave. The latter
gives a standard surface wave with a constant amplitude, which cannot
accept the power of the illuminating plane wave.}.

\begin{figure}[h]
	\centering
        \includegraphics[width=0.25\textwidth]{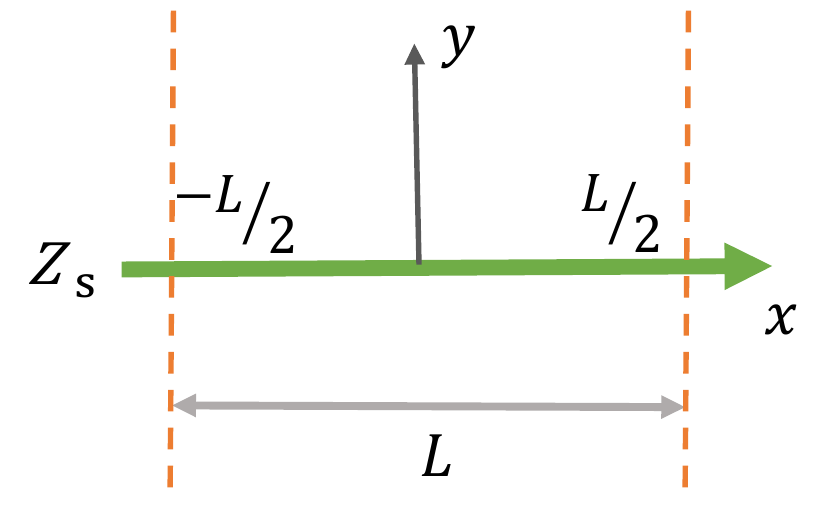}
        \caption{Illustration of the finite interval of the impenetrable surface with length $L$. }
        \label{fig:2}
\end{figure}

Since an exponential function either diverges or
approaches zero as $x\to\pm\infty$, we can only consider approximate
satisfaction of (\ref{cond:hsw0mag}) over a finite interval.
Consider an interval of length $L$ centered at $x=0$, as shown in
Fig.~\ref{fig:2}. We can select the value of $\alpha_x$ such that
$e^{\alpha_xx}$ can be considered approximately equal to unity over
$-L/2<x<L/2$. In other words, attenuating constant 
$\alpha_x$ and length $L$ should be balanced to satisfy the 
condition $|\alpha_xL|\ll 1$.
Picking the middle point of the $x$-range at $x=0$
for exact enforcement of (\ref{cond:hsw0mag}) and setting the phase of
$H^\sw_0$ equal to that of $E_0$, we assign
\begin{equation}
|H^\sw_0|=\frac{E_0}{\eta_0}\sqrt{\frac{k_0}{\beta_y}}.
\label{set:hsw0}
\end{equation}

The $x$-component of the time-average Poynting vector of the surface wave
is equal to
\begin{equation}
S_x=\frac{1}{2}\text{Re}\{E^\sw_y\times H^{\sw*}_z\}
=\frac{\eta_0\beta_x}{2k_0}|H^\sw_0|^2e^{-2(\alpha_xx+\alpha_yy)}.
\label{sswx}
\end{equation}
Integrating $S_x(x,y)$ over $0<y<\infty$, we find that the power
propagating in the $+x$-direction per unit length in $z$ to be
\begin{equation}
P_x=\frac{\eta_0\beta_x}{4k_0\alpha_y}|H^\sw_0|^2e^{-2\alpha_xx}
\approx\frac{\eta_0\beta_x}{4k_0\alpha_y}|H^\sw_0|^2(1-2\alpha_xx).
\label{pswx}
\end{equation}
Equation~(\ref{pswx}) shows that the power carried by the surface wave
increases approximately linearly with $x$, as desired. In addition,
the rate of power increase with respect to 
$x$---$dP_x/dx$---is equal
to the power density of the incident plane wave.

A surface of finite length $L$ is expected to convert the
power of the illuminating plane wave falling on the given range,
$-L/2<x<L/2$, into a growing surface wave propagating in the
$+x$-axis direction. We note that the surface wave field values
at the starting point of the surface $(x=-L/2)$ is not zero.
This means that the wave-converting surface needs an input power
at $x=-L/2$ in an eigenmode characterized by a damped oscillatory
function in $y$ to perform the desired wave conversion, but there is no input power if the considered section of length $L$ has  no continuation at $x<-L/2$.
For this reason, for numerical
validation of the designed surfaces in Section~\ref{sec:results},
performance of the wave converting surfaces is evaluated without
an input surface wave as well.

\section{Surface Impedance Characterization of the Coupler}
\label{sec:4}

To ascertain which type of a surface is needed for 
the transformation described in the previous section,
we consider the boundary condition at $y=0$
in the form of a surface impedance. 
Using superposition,
the total tangential fields
on the surface
are found from (\ref{def:hsw})--(\ref{def:esw}) and
(\ref{def:ei_hi}) to be

\begin{eqnarray}
\bE_{\rm t}&=& \zhat E_{{\rm t}z}+\xhat E_{{\rm t}x}= 
\zhat E_0-\xhat H^\sw_0\frac{k_y\eta_0}{k_0}e^{-jk_xx},
\label{def:Et}\\
\bH_{\rm t}&=&\zhat H_{{\rm t}z}+\xhat H_{{\rm t}x}
=\zhat H^{\rm sw}_0e^{-jk_xx}-\xhat\frac{E_0}{\eta_0},
\label{def:Ht}
\end{eqnarray}
where $E_{{\rm t}z}$ and $E_{{\rm t}x}$ are the
tangential electric field components derived from 
the incident and reflected (quasi-surface) waves.
The magnetic field components $H_{{\rm t}x}$ 
and $H_{{\rm t}z}$ are denoted similarly.
Now, the $y=0$ surface can be characterized with a surface
impedance tensor $\=Z_\text{s}$, which relates $\bE_\text{t}$
and the induced surface current $\bJ_\text{s}=\yhat\times\bH_\text{t}$
via
\begin{equation} 
\bE_\text{t}=\=Z_\text{s}\cdot \bJ_\text{s}=
\begin{bmatrix} 
    Z_{xx} & Z_{xz}\\
    Z_{zx} & Z_{zz}
\end{bmatrix}
\bJ_\text{s},
\label{eq:tensor_bc}
\end{equation}
where $Z_{xx}$, $Z_{xz}$, $Z_{zx}$, and $Z_{zz}$ are terms of 
the anisotropic $2\times 2$ matrix $\=Z_{\rm s}$.
The two complex-valued equations in terms of
field quantities are
\begin{equation}
\begin{array}{l}\displaystyle
E_{{\rm t}x}=Z_{xx}H_{{\rm t}z}-Z_{xz}H_{{\rm t}x},\\
E_{{\rm t}z}=Z_{zx}H_{{\rm t}z}-Z_{zz}H_{{\rm t}x}.
\label{eq:tang}
\end{array}
\end{equation}
Each of the four matrix elements can be
represented as a complex number, i.e.,
\begin{equation}
\=Z_\text{s}=
\begin{bmatrix} 
R_{xx}+jX_{xx} & R_{xz}+jX_{xz}\\
R_{zx}+jX_{zx} & R_{zz}+jX_{zz}
\end{bmatrix},
\label{eq:Z_in_RX}
\end{equation}
where the $R$ and $X$ quantities are real-valued
and represent the resistance and reactance parts of the associated
impedance values.
Since there are two complex-valued equations in (\ref{eq:tang}) for
four complex-valued impedance elements, 
the solution is not unique and thus
there is a freedom that can be exploited for setting
their values.
{\color{black} Equating the real and imaginary parts on the
two sides of (\ref{eq:tang}), we obtain four real-valued
equations. They can be expressed as a matrix equation in
a compact form as
\begin{widetext}
\begin{equation}
\begin{bmatrix}
R_{xx} & R_{xz}\\
R_{zx} & R_{zz}
\end{bmatrix}
\begin{bmatrix}
{\rm Re}\{H_{\t{t}z}\} & {\rm Im}\{H_{\t{t}z}\}\\
-{\rm Re}\{H_{\t{t}x}\} & -{\rm Im}\{H_{\t{t}x}\}
\end{bmatrix}
+
\begin{bmatrix}
X_{xx} & X_{xz}\\
X_{zx} & X_{zz}
\end{bmatrix}
\begin{bmatrix}
-{\rm Im}\{H_{\t{t}z}\} & {\rm Re}\{H_{\t{t}z}\}\\
{\rm Im}\{H_{\t{t}x}\} & -{\rm Re}\{H_{\t{t}x}\}
\end{bmatrix}
=\begin{bmatrix}
{\rm Re}\{E_{\t{t}x}\} & {\rm Im}\{E_{\t{t}x}\}\\
{\rm Re}\{E_{\t{t}z}\} & {\rm Im}\{E_{\t{t}z}\}
\end{bmatrix}.
\label{eq:bc_compact}
\end{equation}
\end{widetext}
}
Depending on requirements on 
loss and reciprocity of the system,
{\color{black}different constraints can be placed on the values of the
resistance and reactance parts of the tensor elements}
and the {\color{black}associated}
impedance/admittance matrices can be found
{\color{black}from (\ref{eq:bc_compact}).}
A reciprocal system is characterized by a symmetric matrix, i.e.,
$\=Z_\text{s}^T=\=Z_\text{s}$.
As to losses in the system, a {\color{black} point-wise} lossless system is
characterized by a 
skew-Hermitian matrix, i.e.,
$\=Z_{\rm s}^\dag=-\=Z_{\rm s}$.
In the following, we discuss solutions for $\=Z_\text{s}$
for reciprocal/non-reciprocal and
lossless/active-lossy combinations.

\subsection{Reciprocal and point-wise lossless}
\label{subsec:4a}

Let us first describe the most desirable case from the practical point of view, when the system is reciprocal and lossless {\color{black} at every point}. 
All the elements of the impedance matrix are purely imaginary 
and the matrix is skew-Hermitian: 
$Z_{xx}=jX_{xx}$, $Z_{xz}=Z_{zx}=jX_{xz}$, and $Z_{zz}=jX_{zz}$.
{\color{black} There are three real-valued parameters 
($X_{xx}$, $X_{xz}$, and $X_{zz}$), while there are 
four linear equations in (\ref{eq:bc_compact}).}
 
\subsubsection{\color{black} Periodic approximation}
\label{subsubsec:periodic}
{\color{black} One can solve all four equations exactly in the case of magnetic field magnitude described by (\ref{cond:hsw0mag}).}
The solutions for the impedance matrix as well as the associated admittance matrix are found to be
\begin{gather}
\=Z_{\rm s}=j
\begin{bmatrix}
\displaystyle\frac{\eta_0}{k_0}\left(\alpha_y-\beta_y\cot\beta_xx\right) & 
\displaystyle\frac{\eta_0}{\sin\beta_xx}\sqrt{\frac{\beta_y}{k_0}}\\
\displaystyle\frac{\eta_0}{\sin\beta_xx}\sqrt{\frac{\beta_y}{k_0}} &
-\eta_0\cot\beta_xx
\end{bmatrix},
\label{eq:Zs1}\\
\=Y_{\rm s}=\frac{j}{D}
\begin{bmatrix}
-\eta_0\cot\beta_xx & \displaystyle-\frac{\eta_0}{\sin\beta_xx}\sqrt{\frac{\beta_y}{k_0}}\\
\displaystyle-\frac{\eta_0}{\sin\beta_xx}\sqrt{\frac{\beta_y}{k_0}} &
\displaystyle\frac{\eta_0}{k_0}\left(\alpha_y-\beta_y\cot\beta_xx\right)
\end{bmatrix},
\label{eq:Ys1}
\end{gather}
where $D$ is the determinant of $\=Z_\text{s}$ 
and is equal to $D=\eta_0^2\left(\alpha_y\cot\beta_xx+\beta_y\right)/k_0$.
This solution describes a completely lossless structure, but the fields satisfy Maxwell's equations exactly only when $e^{\alpha_xx}$ approaches unity.
\begin{figure}[t]
	\centering
        \includegraphics[width=0.45\textwidth]{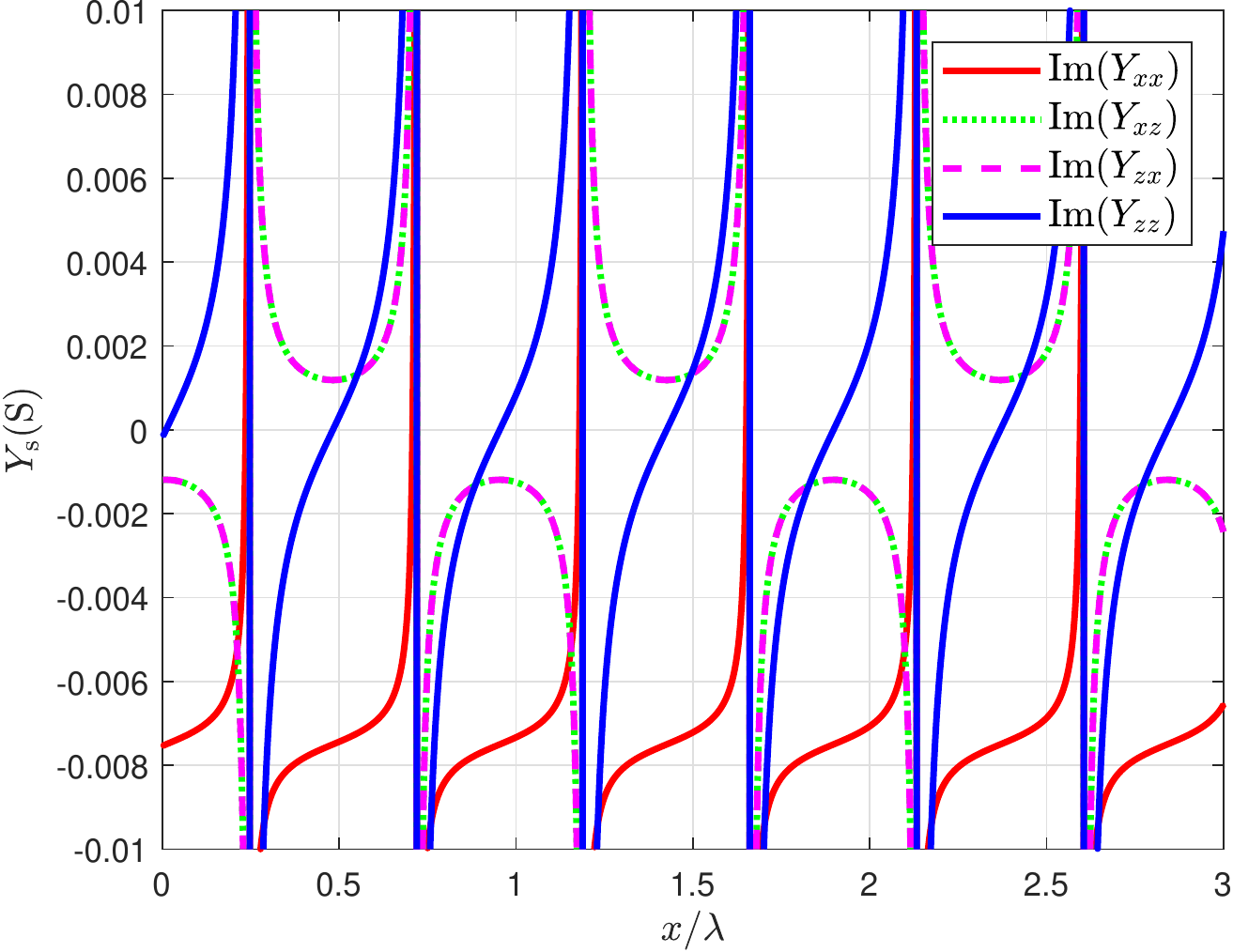}
        \caption{Susceptance elements of the surface admittance $Y_{\rm s}$ for the reciprocal and lossless {\color{black} periodic approximation (\ref{eq:Ys1})} with $\alpha_{x}=-0.0083 k_0$, 
				$\alpha_{y}=0.3524k_0$,  $\beta_{x}=1.06 k_0$, and
				$\beta_{y}=0.025 k_0$.}
        \label{fig:A}
\end{figure}
For an example set of complex propagation constants
(their values are shown in the caption),
Fig.~\ref{fig:A} plots the susceptance elements of
$\=Y_\text{s}$ {\color{black} (\ref{eq:Ys1})}.
The propagation constants are chosen for a surface of
length $L=20\lambda_0$. Along the $+x$-axis direction, a moderately slow
exponential growth was chosen with {\color{black}$|\alpha_xL|\ll 1$} and a propagation
constant just outside the visible (propagating)  range was selected as 
$\beta_x=1.06k_0$. The associated values of $\alpha_y$, $\beta_y$
are found from
(\ref{fs_disp_real})--(\ref{fs_disp_imag}). The admittance tensor
parameters are found to be periodic functions {\color{black} with period $2\pi/\beta_x$}. Furthermore, all four
parameters diverge when $D=0$, or at
$x=(1/\beta_x)[-\cot^{-1}(\beta_y/\alpha_y)+n\pi]$
$(n=0,\pm 1,\pm 2,\ldots)$.

\subsubsection{\color{black} Least squares approximation}
\label{subsubsec:leastsq}
\begin{figure}[t]
	\centering
        \includegraphics[width=0.45\textwidth]{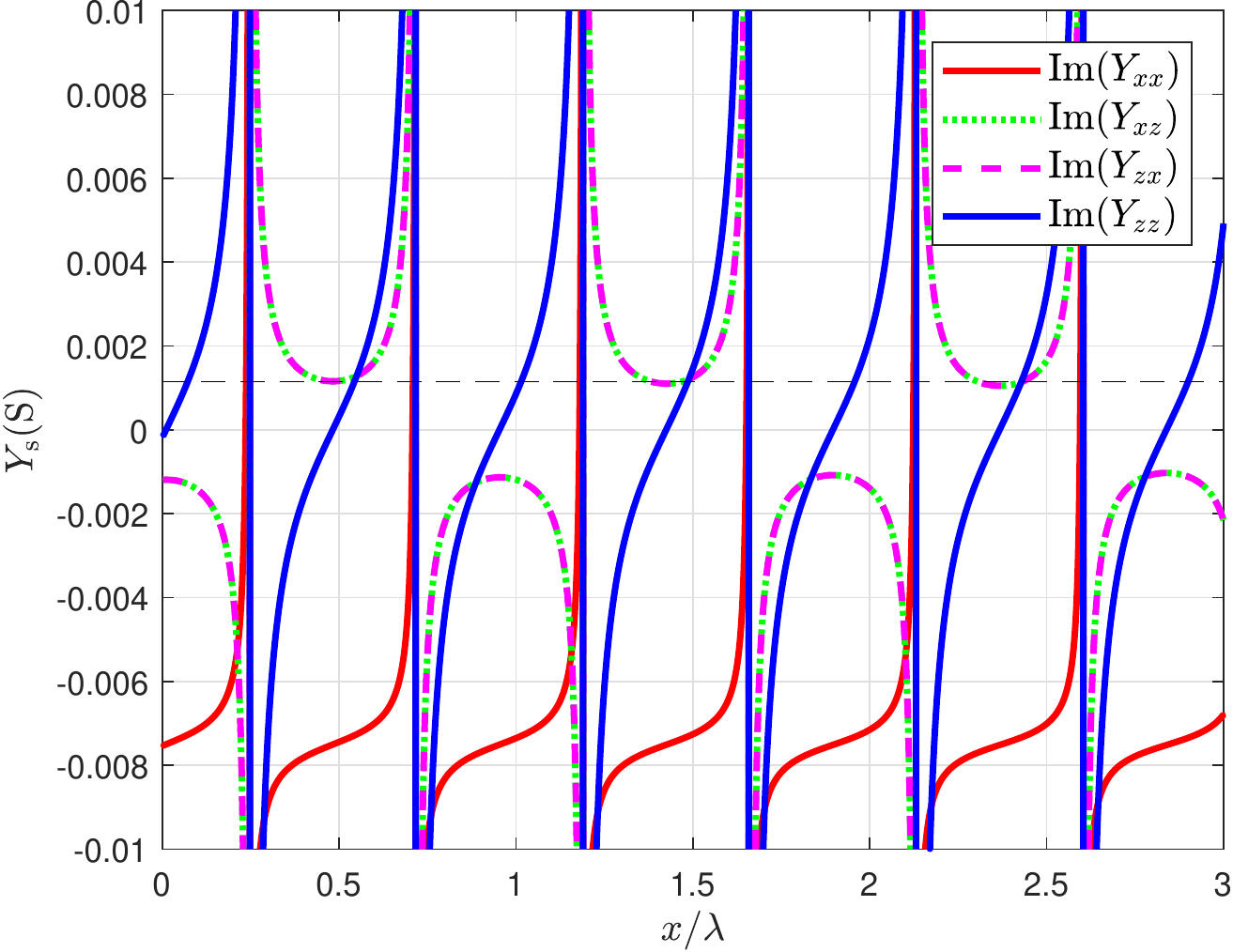}
        \caption{\color{black} Susceptance elements of the surface admittance $Y_{\rm s}$ for the reciprocal and lossless least squares approximation with $\alpha_{x}=-0.0083 k_0$, $\alpha_{y}=0.3524k_0$,  $\beta_{x}=1.06 k_0$, and $\beta_{y}=0.025 k_0$.}
        \label{fig:A3}
\end{figure}

{\color{black}To find an approximation for the  surface impedance/admittance of a lossless and 
reciprocal surface, which creates the fields that satisfy Maxwell's equations, the least squares approximation \cite{strang2016} can be used. The amplitude of the magnetic field of the surface wave in this case is defined as (\ref{set:hsw0})}.  
{\color{black}In terms of the vector with three reactance elements
to be determined, 
$\bX=\begin{bmatrix}
X_{xx} & X_{xz} & X_{zz}
\end{bmatrix}^T$, 
the four linear
equations (\ref{eq:bc_compact}) can be rewritten as
\begin{equation}
\bar{\bar{H}}_\t{t}\bX=\bE_t,
\end{equation}
where
\begin{eqnarray}
\bar{\bar{H}}_{\rm t}&=&\begin{bmatrix}
-{\rm Im}\{H_{{\rm t}z}\} & {\rm Im}\{H_{{\rm t}x}\} & 0\\
{\rm Re}\{H_{{\rm t}z}\} & -{\rm Re}\{H_{{\rm t}x}\} & 0\\
0 & -{\rm Im}\{H_{{\rm t}z}\} & {\rm Im}\{H_{{\rm t}x}\}\\
0 & {\rm Re}\{H_{{\rm t}z}\} & -{\rm Re}\{H_{{\rm t}x}\}
\end{bmatrix},\\
\bE_t &=& \begin{bmatrix}
{\rm Re}\{E_{{\rm t}x}\} & {\rm Im}\{E_{{\rm t}x}\} & 
{\rm Re}\{E_{{\rm t}z}\} & {\rm Im}\{E_{{\rm t}z}\}
\end{bmatrix}^T.
\end{eqnarray}
This overdetermined system cannot be exactly satisfied, other than the case in Sec.~\ref{subsubsec:periodic}.
The least squares solution solves for $\bX$ such that
the error defined by
\begin{equation}
\text{error}=\lVert\bE_t-\bar{\bar{H}}_\t{t}\bX\rVert^2
\label{def:sqerr}
\end{equation}
is minimized. The solution is given by \cite{strang2016}
\begin{equation}
\bX=\left(\bar{\bar{H}}_\t{t}^T\bar{\bar{H}}_\t{t}\right)^{-1}
\bar{\bar{H}}_\t{t}^T\bE_t.
\label{X:ls}
\end{equation}
The resulting} admittance profile in this case is generally aperiodic, and the susceptance elements of $\=Y_\text{s}$ are shown in Figure~\ref{fig:A3}. 
The behavior of the admittance curves in both approximations, periodic and least squares, are similar, but differences become 
pronounced 
at locations far from $x=0$. 

An alternative approximate solution which introduces non-reciprocity while the system remains lossless at all points can be found in a similar way, but it does not offer any advantages in possible realizations. 

\subsection{Reciprocal and active/lossy}
\label{subsec:4d}
A possible way to find an exact solution which satisfies Maxwell's equations is to mitigate the condition for absence of losses or gain everywhere at the metasurface plane.  In other words, strong non-local response in metasurface is allowed. One of the reciprocal and active/lossy
solutions which satisfy (\ref{set:hsw0}) {\color{black} identically at all points of the metasurface} is described by the impedance matrix with imaginary diagonal elements elements and {\color{black}complex} off-diagonal elements. 
The impedance matrix and the associated admittance matrix 
are found to be
{\color{black}
\begin{gather}
\=Z_{\rm s}=\begin{bmatrix}
    j \displaystyle{X_{xx}} & \hspace{0.5cm} \displaystyle{R_{xz}+j X_{xz}}  \\
    \displaystyle{R_{xz}+j X_{xz}}  & \displaystyle{j X_{zz}}
\end{bmatrix},
\label{eq:Zs4}\\
\=Y_{\rm s}=\displaystyle {1\over D } \begin{bmatrix}
   j \displaystyle{X_{zz}}  & -(\displaystyle{R_{xz}+jX_{xz}} ) \\
 -(\displaystyle{R_{xz}+jX_{xz}})  & \hspace{0.5cm} j \displaystyle{X_{xx}}
\end{bmatrix},
\label{eq:Ys4}
\end{gather}
where 
\begin{gather}
X_{xx}={\eta_0 \over k_0} \, \left(\alpha_y-\beta_y \cot\beta_x x \right)+{\eta_0 \beta_y \left(1-e^{2 \alpha_x x}\right)\over2 k_0 \cos{\beta_x x} \sin{\beta_x x}},\\
R_{xz}={\eta_0 \over 2 \cos\beta_x x }\sqrt{\beta_y \over k_0} \left(e^{ \alpha_x x}-e^{- \alpha_x x}\right),\\
X_{xz}={\eta_0 \over 2 \sin{\beta_x x }}\sqrt{\beta_y \over k_0} \left(e^{ \alpha_x x}+e^{- \alpha_x x}\right),\\
X_{zz}={\eta_0 \over 2}\left( \tan{\beta_x x}-\cot{\beta_x x}-{e^{-2\alpha_x x} \over \cos{\beta_x x} \sin{\beta_x x}}\right),
\end{gather}
and $D=-X_{xx} X_{zz}-(R_{xz}+jX_{xz})^2$} is the determinant of $\=Z_\text{s}$.

\begin{figure}[t]
	\centering
        \includegraphics[width=0.45\textwidth]{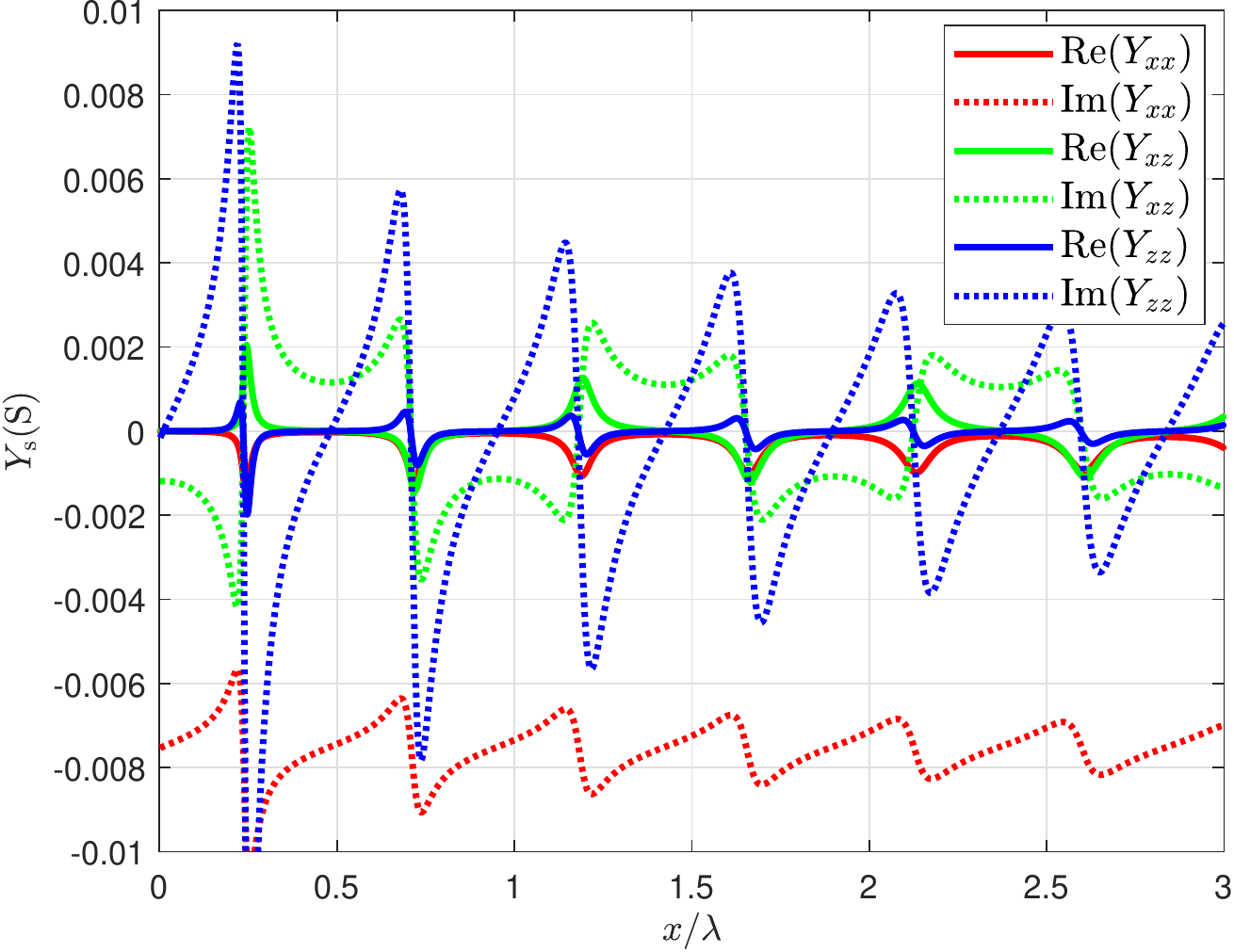}
        \caption{Surface admittance $Y_{\rm s}$ for the reciprocal 
				and active/lossy case with $\alpha_{x}=-0.0083 k_0$, 
				$\alpha_{y}=0.3524k_0$,  $\beta_{x}=1.06 k_0$, 
				$\beta_{y}=0.025 k_0$. {\color{black}All the components of the admittance are complex numbers}.				}
        \label{fig:D}
\end{figure}

Figure~\ref{fig:D} plots the elements of $\=Y_\text{s}$ for 
the same example set of complex propagation constants
considered in previous subsections. {\color{black}All the admittance terms are complex in contrary to the impedance terms due to the fact that the determinant $D$ is  a complex number.}
The admittance tensor parameters are found to be aperiodic functions.

{\color{black}
This revealed possibility to find an exact solution in form of only one incident propagating plane wave 
and one surface mode (with the amplitude given by (\ref{set:hsw0})) using active/lossy metasurfaces reminds of the analogous property of perfect anomalous reflectors which transform one incident plane wave into only one reflected plane wave propagating in the desired direction~\cite{asadchy_prb2016,estakhri_prx2016,epstein_prl2016,diaz-rubio_sciadv2017}. We expect that also in the case of considered transformation into a single \emph{surface} mode, realizations can be found in form of non-local metasurfaces, where ``active'' regions work as receiving leaky-wave antennas while the ``lossy'' regions as transmitting antennas, in analogy with the approach presented in \cite{diaz-rubio_sciadv2017}. In contrast to transformers of propagating waves which are periodically modulated surfaces, in this case the receiving surface is divided into two regions, the first of which is receiving power (effective loss) and the second one is radiating power (effective gain), as is seen from formula (\ref{cond_sy}) with a constant value of $|H^{\rm sw}_0|$. The parameters of the metasurface can be chosen so that these two powers are equal, so that overall  the structure is lossless. } 

{\color{black} It is stressed that realizations of such non-local metasurfaces which emulate active/lossy response do not require active elements. Basically, the non-local metasurface does not act as a boundary at every point: In the ``lossy''  region part of the input power is accepted by auxiliary waves which exist inside the metasurface device, and this power is used to enhance the generated plane wave in the ``active'' region.  Actually, any reactive impedance boundary (except perfect electric conductor)  models some fields behind the surface, only in the point-wise lossless case all the power which is accepted at a given point is reflected back at the same point, without any power transport along the metasurface plane.  In the non-local scenario, some power also moves inside the metasurface in the direction of power growth of the generated plane wave.}

{\color{black}
\subsection{Non-reciprocal and locally active/lossy}
\label{subsec:4b}
 By introducing non-reciprocity in addition to {\color{black} effective} loss and gain, one can find more elegant exact solutions  which satisfy Maxwell's equations, corresponding to creation of a single surface wave with the amplitude  (\ref{set:hsw0})}. It can be described by a non-Hermitian impedance matrix with all the elements 
being purely imaginary: 
$Z_{ xx}=j X_{ xx}$, $Z_{ xz}=j X_{ xz}$,
$Z_{ zx}=j X_{ zx}$, $Z_{ zz}=j X_{ zz}$.
The solution for the impedance matrix as well as the
associated admittance matrix are found to be
\begin{gather}
\=Z_{\rm s}=j\begin{bmatrix}
\centering 
    \displaystyle{\eta_0 \over k_0}\left(\alpha_{y}-\beta_{y} \cot{\beta_{x} x}\right) & \hspace{0.5cm}\displaystyle{\eta_0 \over \sin{\beta_{x}x}}{ \displaystyle\sqrt{\beta_{y} \over k_0  } } e^{-\alpha_{x} x}\\
    \displaystyle{\eta_0 \over \sin{\beta_{x}x}}{ \displaystyle\sqrt{\beta_{y} \over k_0  } } e^{\alpha_{x} x} & -\eta_0 \cot{\beta_{x} x} \end{bmatrix},
\label{eq:Zs2}\\
   \=Y_{\rm s}=\displaystyle {j\over D }\begin{bmatrix}
   -\eta_0 \cot{\beta_{x} x}  & \displaystyle-{\eta_0 \over \sin{\beta_{x}x}}{ \displaystyle\sqrt{\beta_{y} \over k_0  } } e^{-\alpha_{x} x}\\
    \displaystyle-{\eta_0 \over \sin{\beta_{x}x}}{ \displaystyle\sqrt{\beta_{y} \over k_0  } } e^{\alpha_{x} x} & \hspace{0.5cm}\displaystyle{\eta_0 \over k_0}\left(\alpha_{y}-\beta_{y} \cot{\beta_{x} x}\right)
\end{bmatrix},
\label{eq:Ys2}
\end{gather}
where $D$ is the determinant of $\=Z_\text{s}$ and is equal to
$D=\eta_0^2\left(\alpha_y\cot\beta_xx+\beta_y\right)/k_0$.

\begin{figure}[t]
	\centering
        \includegraphics[width=0.45\textwidth]{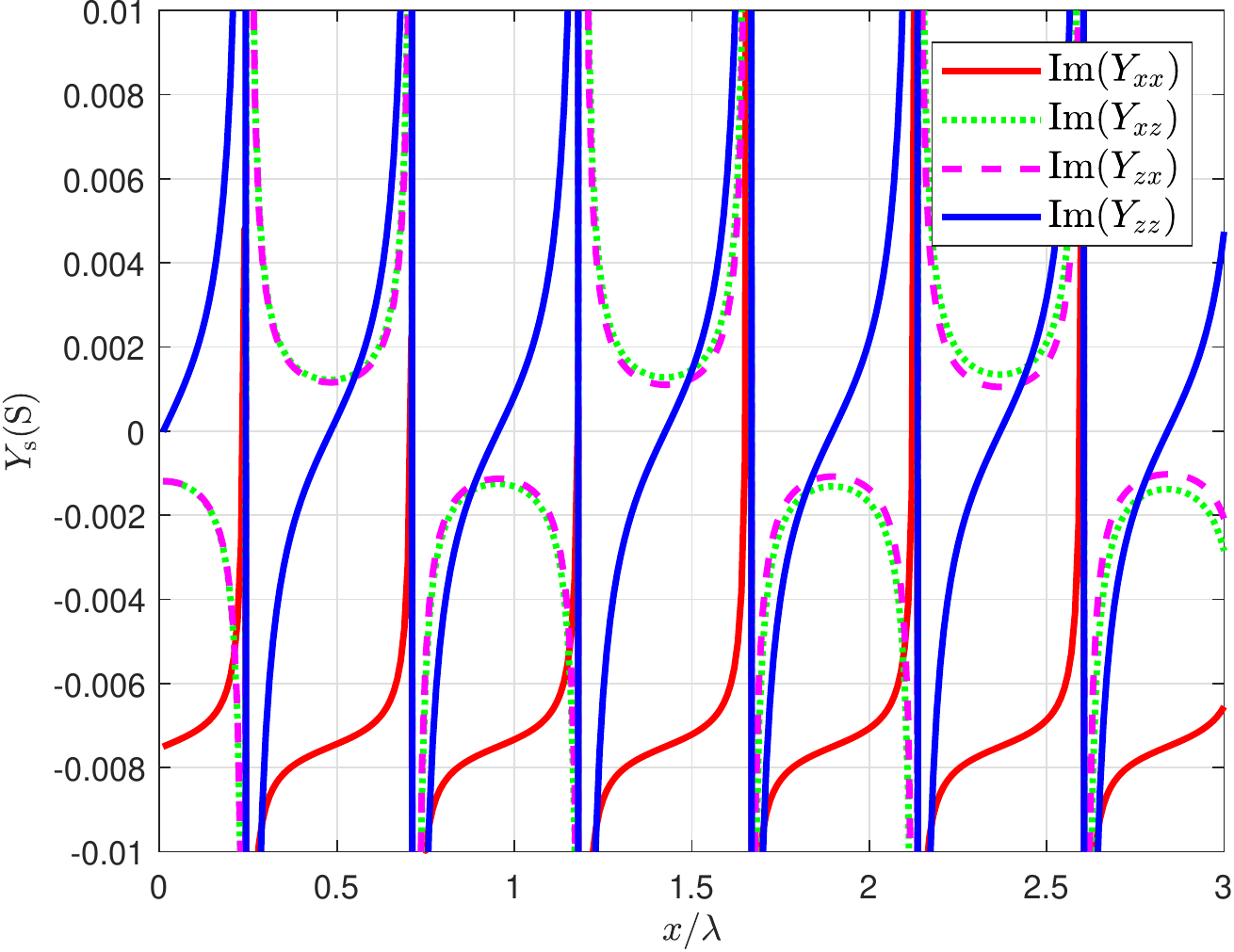}
        \caption{Surface admittance $Y_{\rm s}$ for the non-reciprocal 
				and active/lossy case with $\alpha_{x}=-0.0083 k_0$, 
				$\alpha_{y}=0.3524k_0$,  $\beta_{x}=1.06 k_0$, 
				$\beta_{y}=0.025 k_0$. All the components are purely imaginary.}
        \label{fig:B}
\end{figure}

Figure~\ref{fig:B} plots the elements of the susceptance elements of
$\=Y_\text{s}$ for the same example set of complex propagation constants
considered in Sec.~\ref{subsec:4a}.
The admittance tensor parameters are found to be aperiodic 
functions and all four
parameters diverge when $D=0$, or at
$x=(1/\beta_x)[-\cot^{-1}(\beta_y/\alpha_y)+n\pi]$
$(n=0,\pm 1,\pm 2,\ldots)$.
It is observed that the off-diagonal terms are the
same at $x=0$ and they slowly diverge from each other away from $x=0$.
{\color{black}Therefore, realization of such surface is expected to be
a difficult and probably non-profitable task, compared to reciprocal structures.}
However, they remain close to each other in the $x$-range considered,
signifying that the degree of non-reciprocity and active/lossy property
is not significant. Also as expected, the parameters in
Fig.~\ref{fig:B} are close to those of the reciprocal and lossless
case shown in Fig.~\ref{fig:A}.

By using the aforementioned surface impedance profiles, 
one can simulate and, further, realize the necessary surface to convert 
a propagating plane wave into a quasi-surface wave with high efficiency.

\section{Numerical results}
\label{sec:results}

Using full-wave simulations 
{\color{black}with COMSOL Multiphysics}~\cite{comsol},
the conversion characteristics of the surfaces in
Sec.~\ref{sec:4} can be evaluated.
The model is a rectangular cross section in the
$xy$-plane with a length $L=20 \lambda_0$ and a height
$H=5\lambda_0$. Impedance boundary conditions are 
{\color{black}impenetrable} and applied to the surface
by impressing an electric surface current specified in terms of the tangential electric field and the surface admittance matrix (in our case 
$J_{x}=Y_{xx} E_{{\rm t}x}+Y_{xz} E_{{\rm t}z}$ and 
$J_{z}=Y_{zx} E_{{\rm t}x}+Y_{zz} E_{{\rm t}z}$). 
In this work we consider two cases: imitation of an infinitely long surface with an input surface wave using two surface-wave ports and a model of the initial portion of the wave-converting surface with no input surface-wave port (i.e., one surface-wave port for the converted output power).
The used frequency is $f=10$ GHz, but the results
are scalable to any frequency.

\subsection{Model with an input surface wave} 
\label{subsec:5a}

\begin{figure}[htp]
	\centering
		\begin{gather*}{
        \includegraphics[width=0.35\textwidth]{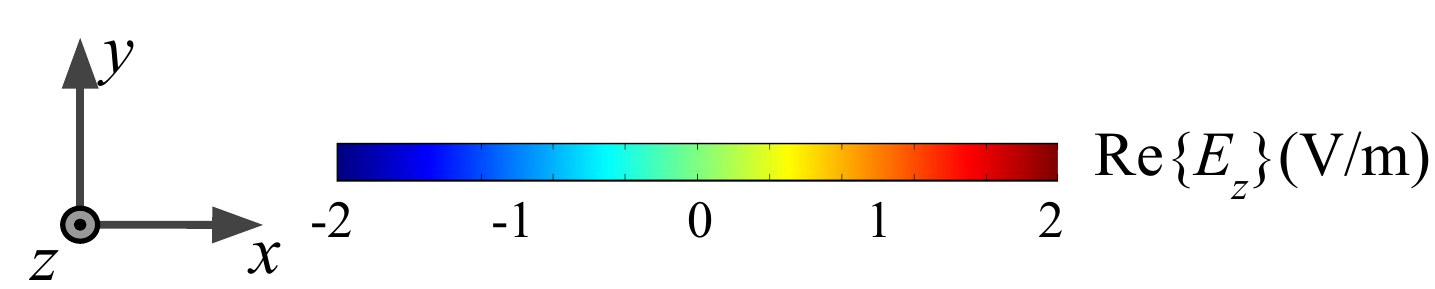}}               
              \end{gather*}  
    \begin{subfigure}[h]{0.45\textwidth}
        \includegraphics[width=\textwidth]{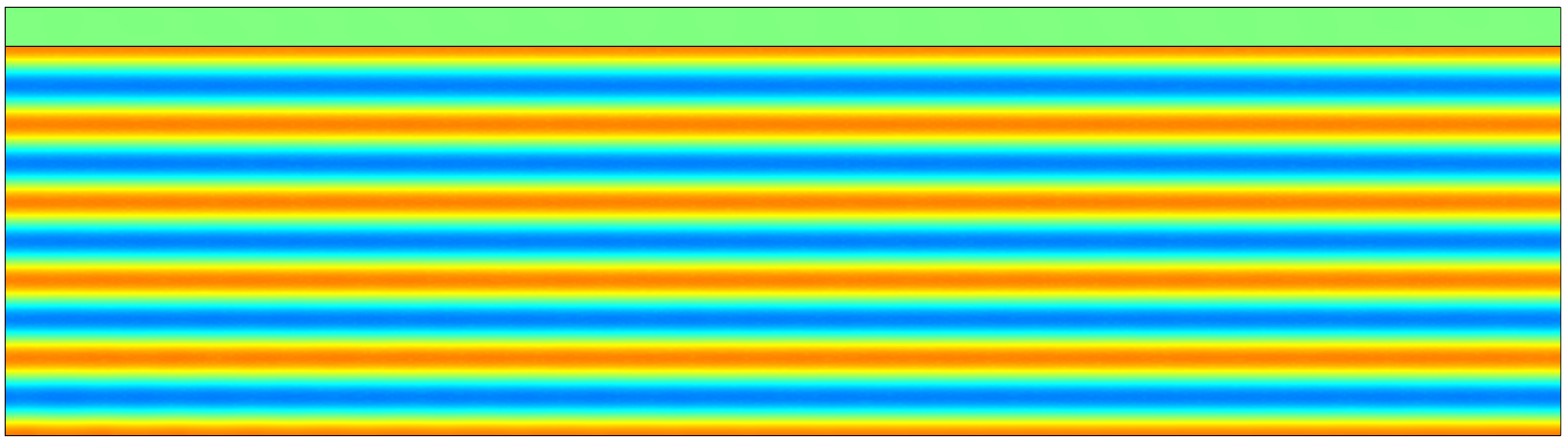}
        \caption{}
    \end{subfigure}\\           
		\begin{gather*}{
        \includegraphics[width=0.35\textwidth]{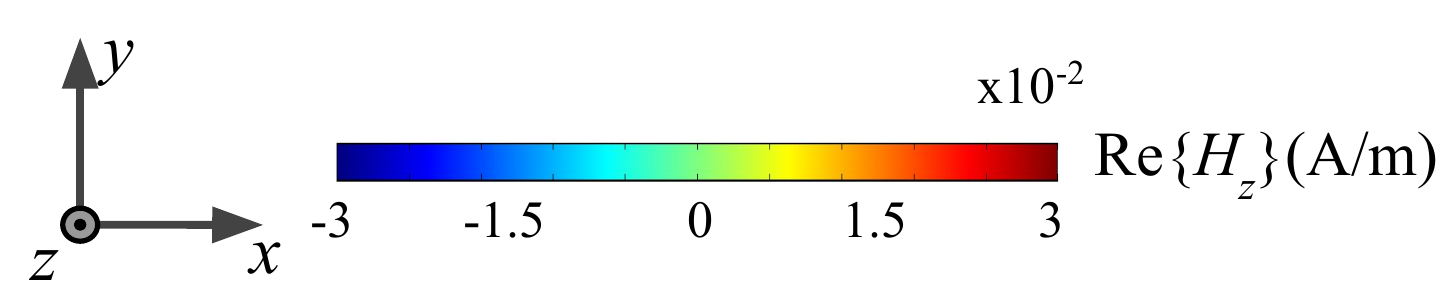}}  
              \end{gather*}
    \begin{subfigure}[h]{0.45\textwidth}
        \includegraphics[width=\textwidth]{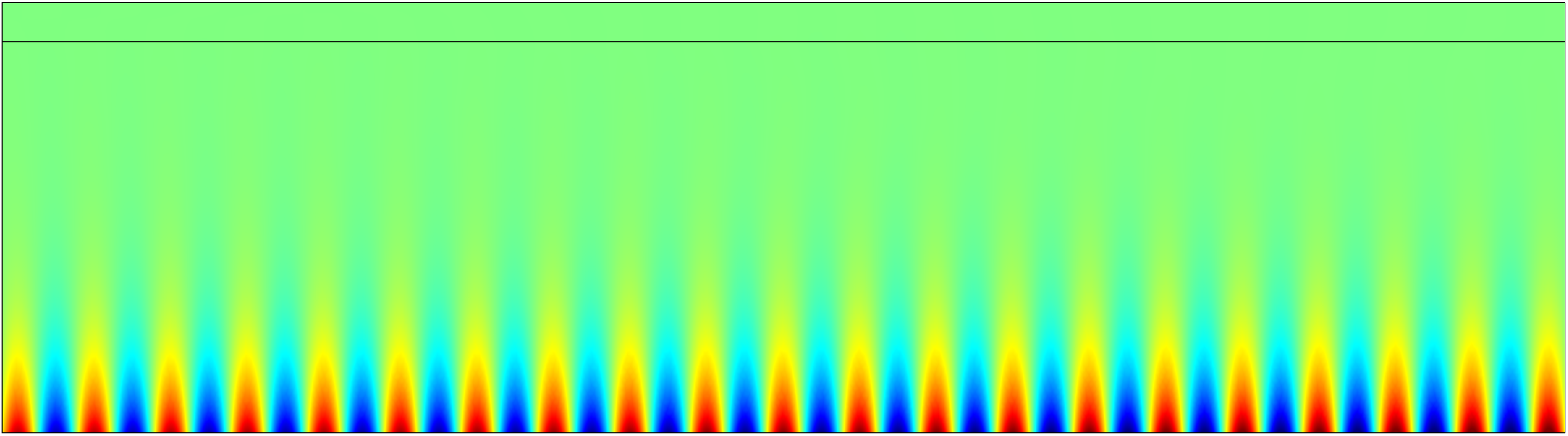}
          \caption{}
            \end{subfigure}\\
    \begin{subfigure}[h!]{0.45\textwidth}
        \includegraphics[width=\textwidth]{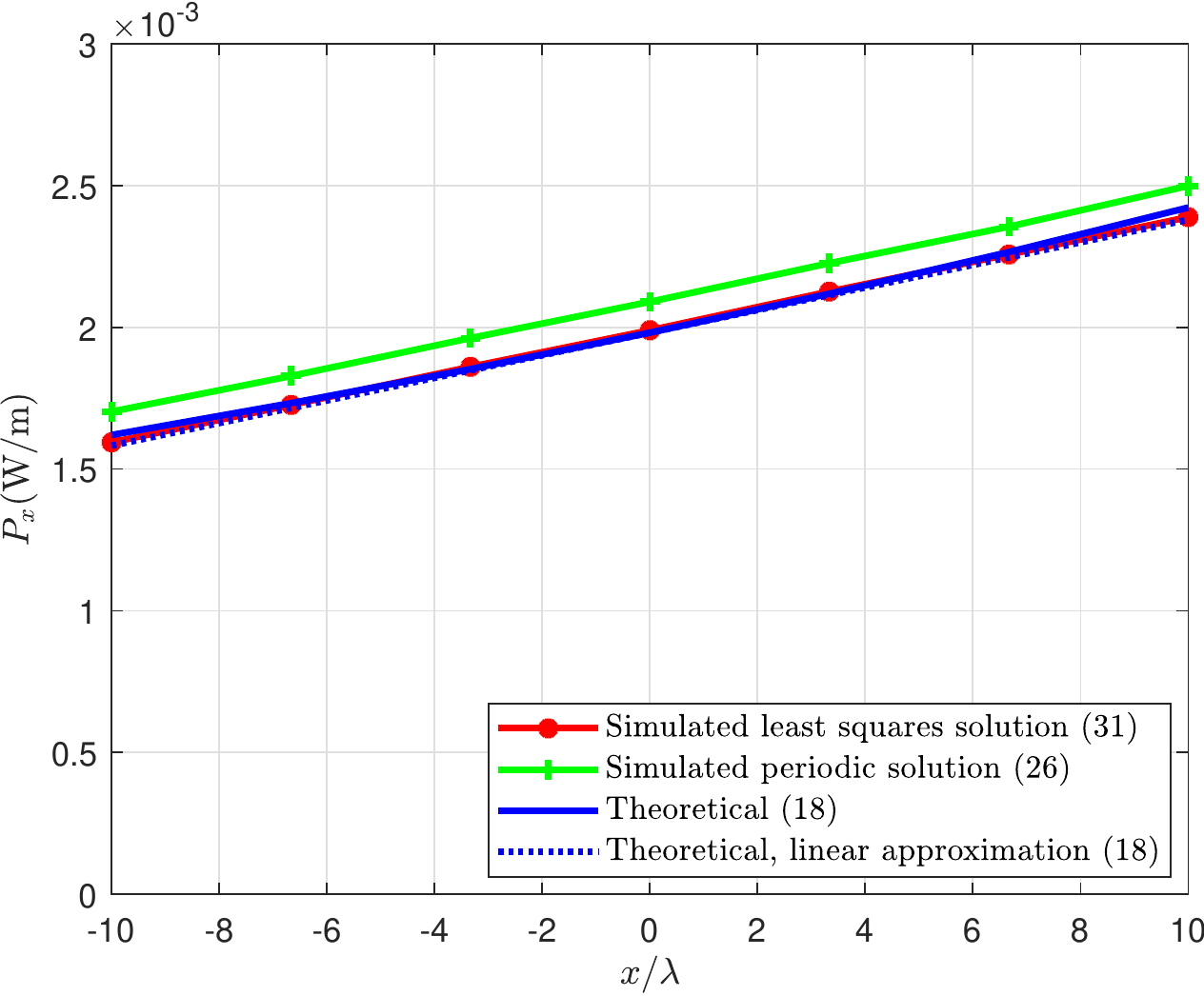}
                \caption{}
    \end{subfigure}
     \caption{
	  (a)  Field distribution of the tangential component of the total field, $E_z$, in the lossless and reciprocal design (least squares solution). The length of the surface is $L=20 \lambda_0$, wave parameters are: $\alpha_{x}=-0.0016 k_0$, $\alpha_{y}=0.2041 k_0$,  $\beta_{x}=1.0206 k_0$, $\beta_{y}=0.008 k_0$.
     (b) Simulated $H_z$ field distribution in the lossless and reciprocal design (least squares solution). Conversion efficiency is nearly $100\%$.
     (c) The tangential power growth along the surface for the 
     least squares {\color{black}(efficiency is 
99.8\%)}, 
     approximate periodic {\color{black}(efficiency is 99.6\%)}, 
     theoretical exponential and theoretical linear approximate solutions.}
            \label{fig:Hinf}
\end{figure}

To model an $L$-long section of the infinite 
structure, port conditions on both sides of the box were defined.
In COMSOL, the mode fields are specified to be
TM-polarized and have a $y$-dependence of $e^{-(\alpha_y+j\beta_y)y}$
on the port surfaces.
Port 1 on the left side is a negative resistor and it pumps energy 
to the system 
to mimic the surface wave that comes from the ``hidden'' part of 
the infinite surface.
This is implemented by setting the propagation constant
normal to the port surface to $-(\beta_x-j\alpha_x)$.
Port 2 on the right side is completely 
passive and receives all the energy carried by a quasi-surface wave.
The conversion efficiency in the case with an input power is 
the fraction of power carried by the surface wave that exits 
$x=L/2$ less the surface wave power that enters $x=-L/2$,
divided by the incident power that falls 
on $-L/2<x<L/2$. 

\begin{figure}[htp]
	\centering
			\begin{gather*}{
        \includegraphics[width=0.35\textwidth]{Re_Ez.pdf}}               
              \end{gather*}  
    \begin{subfigure}[h]{0.45\textwidth}
        \includegraphics[width=\textwidth]{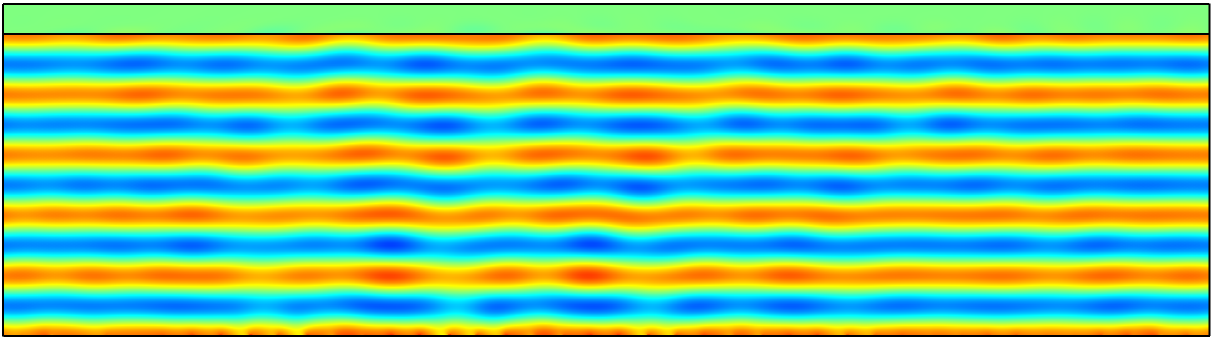}
        \caption{}
    \end{subfigure}\\
		\begin{gather*}{
        \includegraphics[width=0.35\textwidth]{Re_Hz.pdf}}               
              \end{gather*}  
    \begin{subfigure}[h!]{0.45\textwidth}
        \includegraphics[width=\textwidth]{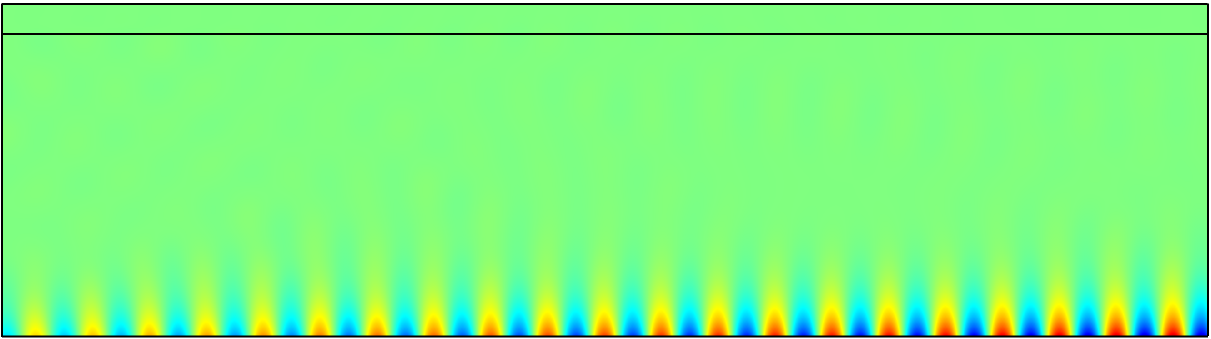}
          \caption{}
            \end{subfigure}\\
    \begin{subfigure}[h!]{0.45\textwidth}
        \includegraphics[width=\textwidth]{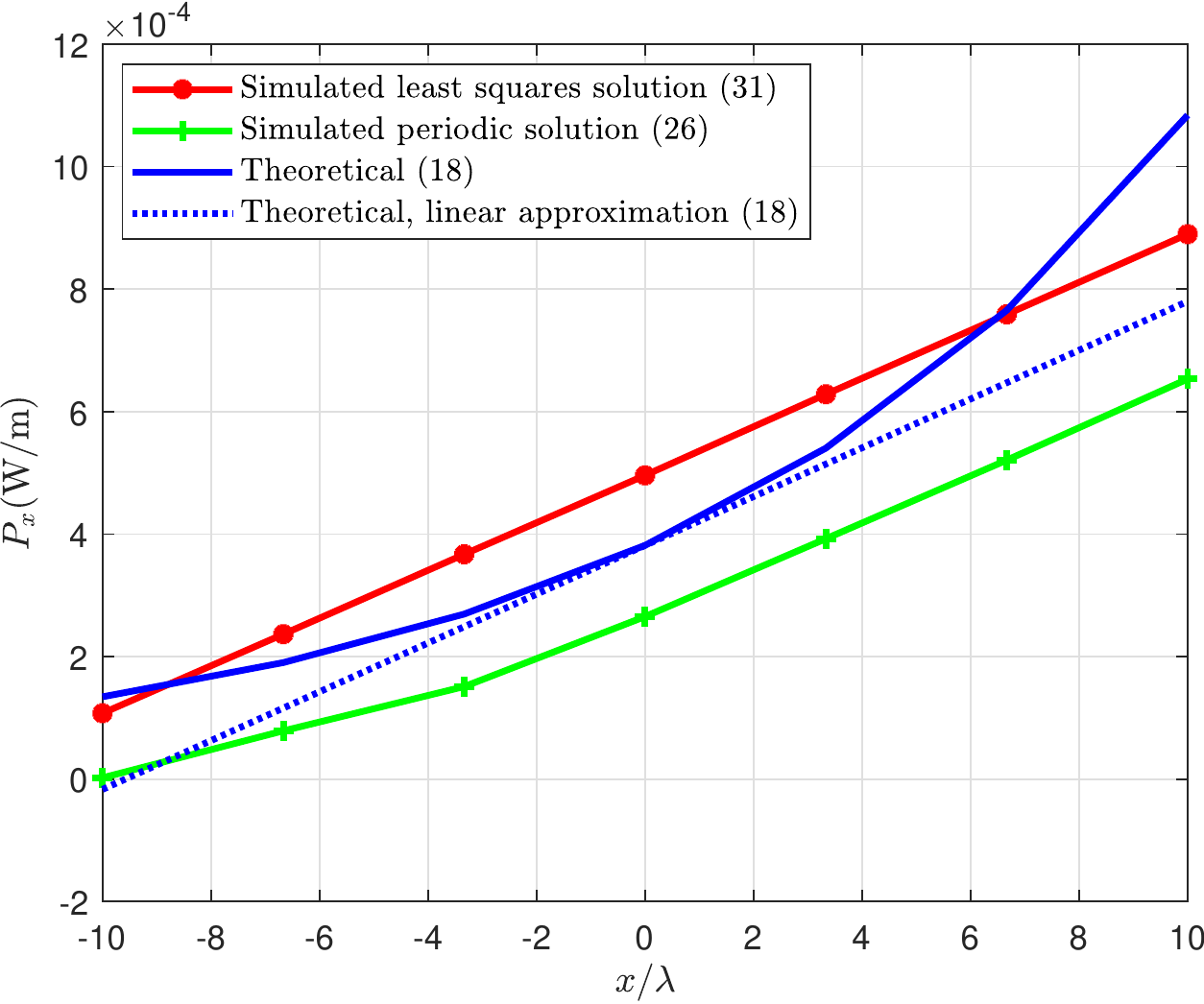}
                \caption{}
    \end{subfigure}
     \caption{ {\color{black}(a)  Field distribution of the tangential component of the total field, $E_z$, in the lossless and reciprocal design (least squares solution). The length of the surface is $L=20 \lambda_0$, wave parameters are: $\alpha_{x}=-0.0083 k_0$, $\alpha_{y}=0.3524 k_0$,  $\beta_{x}=1.06 k_0$, $\beta_{y}=0.025 k_0$.
     (b) Simulated $H_z$ field distribution in the lossless and reciprocal design (least squares solution).
The conversion efficiency is  98.4\%}.
     (c) The tangential power growth along the surface for the 
     least squares {\color{black}(efficiency is 98.4\%)}, 
     approximate periodic {\color{black}(efficiency is 81.7\%), 
     theoretical exponential and theoretical linear approximate solutions.
}}
                 \label{fig:Hinf_2}
\end{figure}

Figure~\ref{fig:Hinf} demonstrates the result for a
lossless and reciprocal system {\color{black}(specifically, 
the least squares solution)}, where the surface 
wave parameters are: $\alpha_{x}=-0.0016 k_0$, 
$\alpha_{y}=0.2041 k_0$,  $\beta_{x}=1.0206 k_0$, and
$\beta_{y}=0.008 k_0$. A snapshot of the $z$-component of the total E-field,
$E_z$, is plotted in Fig.~\ref{fig:Hinf}(a). No reflected propagating
wave is visible and the total field is virtually equal to the
incident field. Figure~\ref{fig:Hinf}(b) plots a snapshot of
$H_z$, showing a slowly growing surface wave that is bound to
the $xz$-plane. Here, $\alpha_{x}$ is a small negative
quantity, therefore the linear approximation of an exponential function is highly accurate and the linear growth of the tangential component of 
the power is smooth. In order to evaluate the wave conversion performance
quantitatively, the total $+x$-directed power per unit length in $z$
[i.e., $P_x$ in (\ref{pswx})] that penetrates an $x=\text{constant}$ plane is shown in
Fig.~\ref{fig:Hinf}(c) with respect to $x$ between theory and
simulation. The simulated power profiles agree well with the linear
approximation of a slow exponential growth.
The efficiency of such conversion is calculated to be 
99.8\% (least squares solution),
which is practically perfect.

Figure~\ref{fig:Hinf_2} demonstrates the result for a
lossless and reciprocal system 
{\color{black}(the least squares solution)} when the surface 
wave parameters are $\alpha_{x}=-0.0083 k_0$, 
$\alpha_{y}=0.3524 k_0$, $\beta_{x}=1.06 k_0$, and 
$\beta_{y}=0.025 k_0$.
A faster exponential growth was chosen compared with
the previous case to investigate the characteristics.
In case of $\alpha_{x}$ being a negative quantity
farther away from zero, the linear approximation is not as
accurate, but the efficiency is still found to be
{\color{black} high at 98.4\%, reduced only slightly from
the previous case in Fig.~\ref{fig:Hinf}.
This slight drop can be understood as a consequence of
a poorer approximation of a faster exponential
growth to a linear growth.}
Reflection of the incident plane wave is minor,
so that $E_z$ is only slightly modified from 
the incident field
{\color{black}[Figure~\ref{fig:Hinf_2}(a)]}.
We also observe that the input power
(normalized to the power density of the plane wave) 
at $x=-L/2$ is significantly
lower than in the small $|\alpha_x|$ case in Fig.~\ref{fig:Hinf}.

Owing to the construction of a growing surface wave solution
at a low exponential rate,
a surface wave input is required for perfect conversion. Although this is an appropriate approach to the analysis of the performance of a middle section of a large receiving surface,  requiring an input wave is not desirable if we consider edges of a finite-length surface.
However, if the required input surface power is low as in
the case shown in
Fig.~\ref{fig:Hinf_2}, it may still be possible to achieve a high
efficiency without an input wave. This possibility is 
investigated next.

\subsection{Model without an input surface wave}
\label{subsec:5b}
From one side we define the ``starting point'' of the surface 
by eliminating the input surface-wave port in previous
configurations and applying perfectly matched layer from the left side 
(the same way we can apply passive port condition, 
their impact here is no different). To the right side we 
connect a port, which receives all the energy 
carried by the quasi-surface wave. The conversion efficiency 
in the case of an absence of the input power is the fraction 
of power carried by the surface wave that exits $x=L/2$ 
relative to the incident power that falls on $-L/2<x<L/2$. 
 It is important to note that the plane-to-surface waves conversion efficiency in the case of no input surface wave is equal to the conversion efficiency of the corresponding leaky-wave antenna \cite{Maci_eff} 
in a reciprocal operation as a radiator. The aperture efficiency equals 100\% by design, and the radiation efficiency approaches 100\% for lossless surfaces.

\begin{figure}[h!]
	\centering
				\begin{gather*}{
        \includegraphics[width=0.35\textwidth]{Re_Ez.pdf}}               
              \end{gather*}  
    \begin{subfigure}[h]{0.45\textwidth}
        \includegraphics[width=\textwidth]{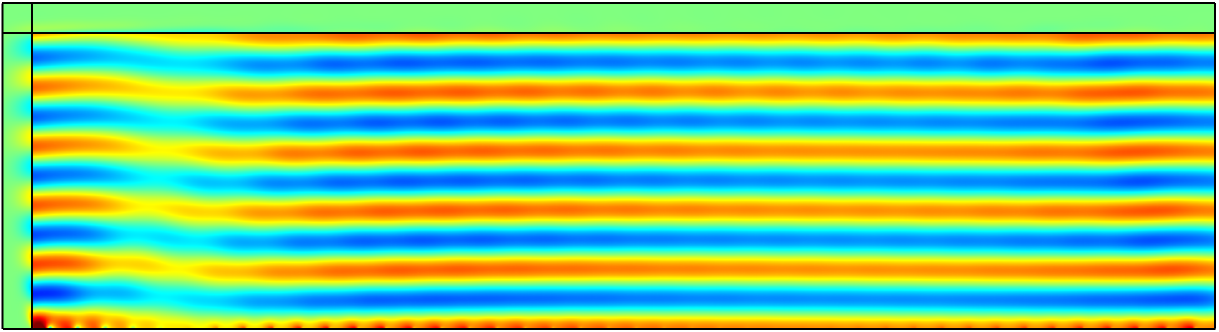}
        \caption{}
    \end{subfigure}\\
		\begin{gather*}{
        \includegraphics[width=0.35\textwidth]{Re_Hz.pdf}}               
              \end{gather*}  
    \begin{subfigure}[h]{0.45\textwidth}
        \includegraphics[width=\textwidth]{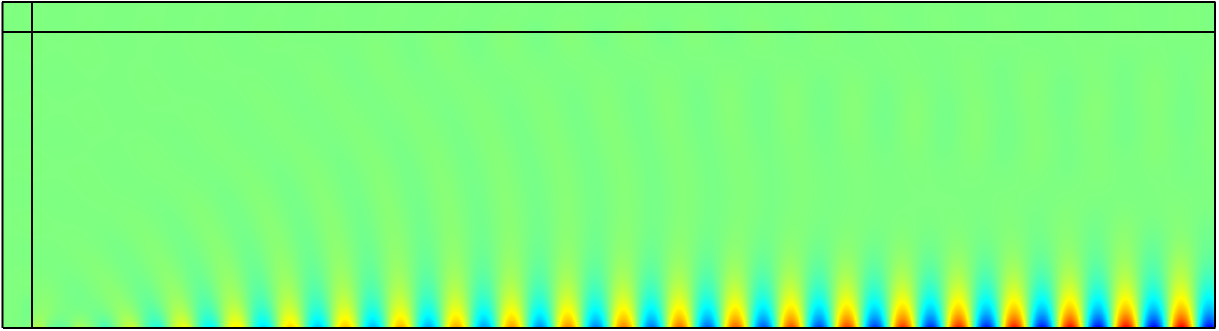}
          \caption{}
            \end{subfigure}\\
         \begin{subfigure}[h]{0.45\textwidth}
        \includegraphics[width=\textwidth]{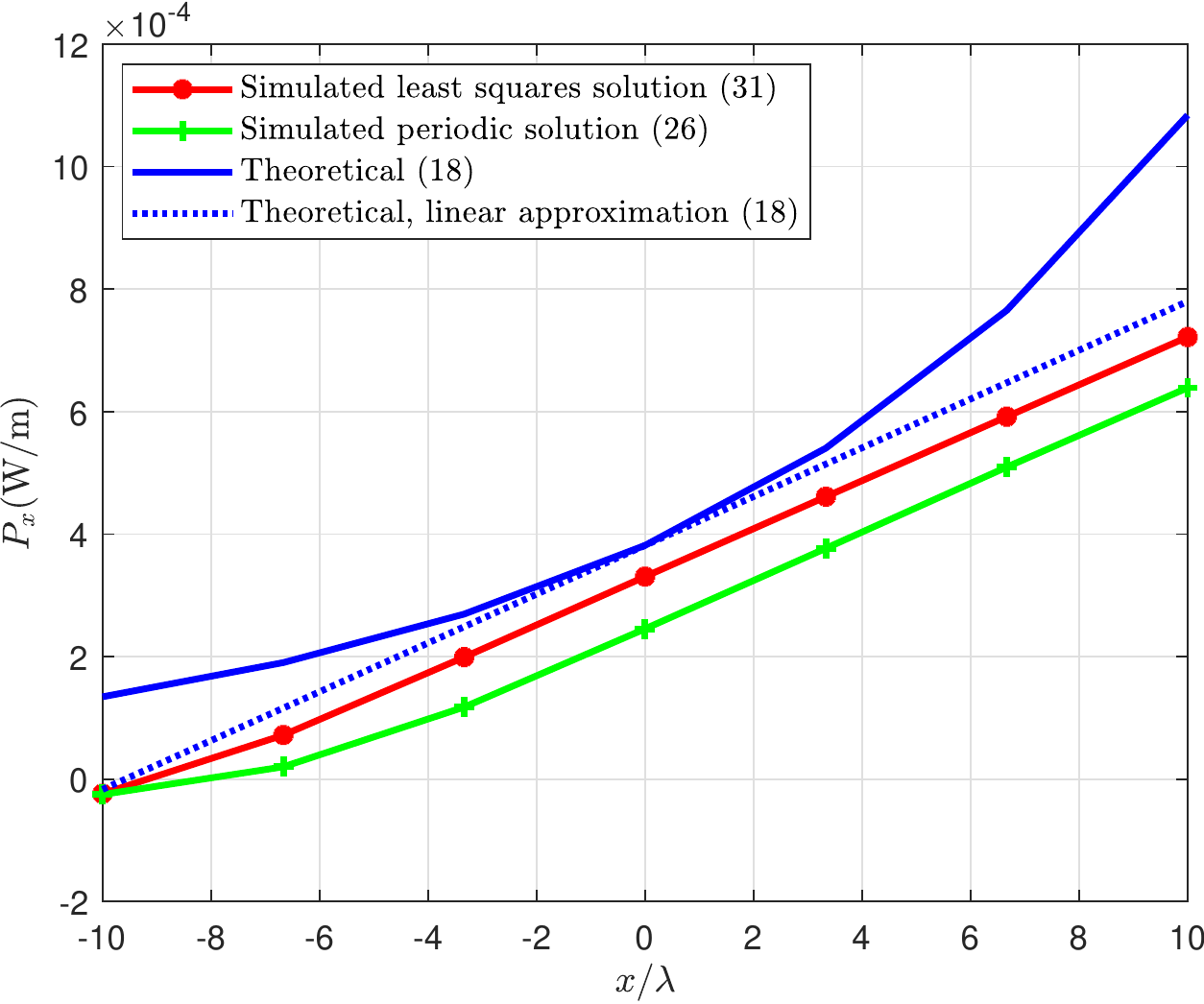}
 \caption{}
    \end{subfigure}
     \caption{ (a)  Field distribution of the tangential component of the total field, $E_z$, in the lossless and reciprocal design (least squares solution). The length of the surface is $L=20 \lambda_0$, wave parameters are: $\alpha_{x}=-0.0083 k_0$, $\alpha_{y}=0.3524 k_0$,  $\beta_{x}=1.06 k_0$, $\beta_{y}=0.025 k_0$.
     (b) Simulated $H_z$ field distribution in the lossless and reciprocal design (least squares solution). The conversion efficiency is 90.5\%. 
     (c) The tangential power growth along the surface for the 
     least squares (efficiency is 90.5\%), 
     approximate periodic (efficiency is 80.2\%), 
     theoretical exponential and theoretical linear approximate solutions. 
}
     \label{fig:Hfin}
\end{figure}

With the same set of surface wave parameters
as in Fig.~\ref{fig:Hinf_2}, snapshots 
of the $z$-component of the total E-field, $E_z$, and the
surface wave magnetic field distribution for the lossless 
and reciprocal surface design (least squares solution)
are shown in Fig.~\ref{fig:Hfin}(a) 
and Fig.~\ref{fig:Hfin}(b), respectively.
The power $P_x$ is plotted in Fig.~\ref{fig:Hfin}(c)
for the analytical and numerical results.
 We note that in numerical simulations the perfectly matched layer on the left side unfortunately distorts 
the incident plane wave, not quite adequately modelling the plane wave.
Moreover, slightly negative values for the simulated power at
$x=-L/2$ are due to the diffracted waves contributing to
power propagating along the  $-x$-direction. However, after a short transition,
a steadily growing surface-wave behavior is established.
Conversion efficiencies for the least squares
and approximate periodic solutions 
are numerically found to be $90.5\%$ and
$80.2\%$, respectively.
These efficiencies are further reduced from the case in Fig.~\ref{fig:Hinf_2} due to the omission of the input surface wave required by the theory. 
This result means that the designed metasurface does not suffer from significant efficiency losses even if used 
without an input power. 
In addition, it is noted that the growing surface wave parameters shown in Fig.~\ref{fig:Hfin} represent one of many design possibilities.
There may be other choices of $\alpha_x$ and $\beta_x$ that lead to higher efficiency values.
Preliminary numerical tests show that efficiency values as high as 95\% are possible for the $20\lambda$-long surface considered in this study.

\section{Conclusions}
\label{sec:conclusion}

In this paper the conversion of a propagating plane wave into a surface wave has been examined theoretically and numerically. In the theoretical discussion the limitations resulting from the required linear growth of the power carried by the surface wave along the direction of propagation are revealed. We have proved that for this spatial power dependence, both separable and non-separable eigenwave field solutions to the Helmholtz equations are unable to represent the surface wave converted from incident propagating plane waves, for any point-wise lossless receiving surface. Next we have shown that a properly constructed  approximate separable solution with a slow exponential growth of the fields can serve as an accurate approximation of the ideally converted surface wave. Moreover, we have proposed several alternative design scenarios, leading to specific surface impedance profiles of nearly ideal wave-converting metasurfaces.

Furthermore, we have shown that dropping the requirement of local (at every point) passivity of the receiving surface potentially opens up possibilities for creation of perfect propagating/surface mode converters using non-local metasurfaces. In this new scenario, the surface first acts as a receiving leaky-wave antenna, emulating an absorbing surface. The received power is then transported along the surface and radiated into space, emulating an active surface. We expect that such locally active/lossy but overall lossless converters can be realized as carefully designed nonuniform patch arrays, generalizing the approach used in \citep{diaz-rubio_sciadv2017} for anomalous reflectors.

Out of the metasurface designs, the tensor surface impedance
based on the least squares solution of the boundary condition and non-local metasurfaces emulating the active/lossy surface impedance allow realization using low-loss, reciprocal constituents.
Arrays of printed subwavelength resonators on a grounded dielectric
substrate~\cite{fong_ieeejap2010,minatti_ieeejap2012,patel_ieeejap2013}
are prime candidates for realizing the required surface reactance tensor.  
In the case of non-local metasurfaces, the required active/lossy profile is realized at some electrically small distance from the patch arrays, where the auxiliary reactive fields effectively decay \cite{diaz-rubio_sciadv2017}. 
The position-dependent shape, size, and rotation angle of an anisotropic printed resonator are determined based on the eigenvalues and eigenvectors of the
reactance tensor. 
Realizing a non-reciprocal surface characteristic will require constituents such as ferromagnetic components~\cite{pozar2005} or magnetless non-reciprocal
devices~\cite{kodera_ieeejmtt2013}.

\section*{Acknowledgement}
This work was supported in part by the Academy of Finland (project 287894) and Nokia Foundation (project 201810155).

\bibliographystyle{apsrev4-1}
\nocite{apsrev41Control}
\bibliography{IEEEabrv,references}

\end{document}